# Thermodynamics and its Prediction and CALPHAD Modeling: Review, State of the Art, and Perspectives

Zi-Kui Liu

Department of Materials Science and Engineering, The Pennsylvania State University, University Park, Pennsylvania 16802, USA

**Abstract:**

Thermodynamics is a science concerning the state of a system, whether it is stable, metastable, or unstable, when interacting with its surroundings. The combined law of thermodynamics derived by Gibbs about 150 years ago laid the foundation of thermodynamics. In Gibbs combined law, the entropy production due to internal processes was not included, and the 2$^{nd}$ law was thus practically removed from the Gibbs combined law, so it is only applicable to systems under equilibrium, thus commonly termed as equilibrium or Gibbs thermodynamics. Gibbs further derived the classical statistical thermodynamics in terms of the probability of configurations in a system in the later 1800's and early 1900's. With the quantum mechanics (QM) developed in 1920's, the QM-based statistical thermodynamics was established and connected to classical statistical thermodynamics at the classical limit as shown by Landau in the 1940's. In 1960's the development of density function theory (DFT) by Kohn and co-workers enabled the QM prediction of properties of the ground state of a system. On the other hand, the entropy production due to internal processes in non-equilibrium systems was studied separately by Onsager in 1930's and Prigogine and co-workers in the 1950's. In 1960's to 1970's the digitization of thermodynamics was developed by Kaufman in the framework of the CALculation of PHAse Diagrams (CALPHAD) modeling of individual phases with internal degrees of freedom. CALPHAD modeling of thermodynamics and atomic transport properties has enabled computational design of complex materials in the last 50 years. Our recently termed zentropy theory integrates DFT and statistical mechanics through the replacement of the internal energy of each individual configuration by its DFT-predicted free energy. The zentropy theory is capable of accurately predicting the free energy of individual phases, transition temperatures and properties of magnetic and ferroelectric materials with free energies of individual configurations solely from DFT-based calculations and without fitting parameters, and is being tested for other phenomena including superconductivity, quantum criticality, and black holes. Those predictions include the singularity at critical points with divergence of physical properties, negative thermal expansion, and the strongly correlated physics. Those individual configurations may thus be considered as the genomic building blocks of individual phases in the spirit of the materials genome®. This has the potential to shift the paradigm of CALPHAD modeling from being heavily dependent on experimental inputs to becoming fully predictive with inputs solely from DFT-based calculations and machine learning models built on those calculations and existing experimental data through newly developed and future open-source tools. Furthermore, through the combined law of thermodynamics including the internal entropy production , it is shown that the kinetic coefficient matrix of independent internal processes is diagonal with respect to the conjugate potentials in the combined law, and the cross phenomena that the phenomenological Onsager flux and reciprocal relationships are due to the dependence of the conjugate potential of a molar quantity on nonconjugate molar quantities and other potentials, which can be predicted by the zentropy theory and CALPHAD modeling.

## 1 Introduction

This paper is based on the author's presentation given at the CALPHAD Global 2021 virtual conference on the perspectives of CALPHAD modeling in next 50 years in view of its success in last 50 years. The future is hard to predict due to its intrinsic uncertainty in terms of probability of many possible events. Nevertheless, it is important to have perspectives on how the future may look like in both short- and long-terms based on past knowledge and anticipated trajectories. In 2020, in celebrating the 50$^{th}$ anniversary of "The Bridge" published by National Academy of Engineering, Sinnott and Liu (the present author) attempted such an effort on "Predicted Advances in the Design of New Materials" [1]. With the prehistory and protohistory of humanity divided into three ages in terms of materials of stone, bronze and iron followed by the three industry revolutions in terms of steam power, electricity, and computerization, we are now entering the 4$^{th}$ industry revolution, i.e., Industry 4.0. Industry 4.0 is commonly thought of as the integration of cyber-physical systems, where the physical, digital, and biological worlds are seamlessly unified to form a system with many autonomous subsystems enabled by advanced materials. In other words, digitization of materials and their manufacturing into functional devices in terms of Materials 4.0 and Manufacturing 4.0 [2]. Such digitization will demand increasingly more efficient development and deployment of materials with emergent properties to meet the performance requirements under extreme conditions. Sinnott and Liu [1] concluded that when this integrated system is fully implemented, the residuals from the design, manufacturing, service, and recycling of materials can be drastically reduced, thus lessening the impact of materials usage on the environment.

In the last century, digitization of materials knowledge progressed significantly, including the digitization of the Schrödinger equation in quantum mechanics [3,4] by the density functional theory (DFT) [5,6], resulting in massive digital databases of material properties predicted using high-performance computers, and thermodynamics by the CALculation of PHAse Diagrams (CALPHAD) method [7–10], resulting in CALPHAD databases widely used in academia and industry for education and design of technologically important materials. Those data together with models and mechanistic correlations are enabling the development of artificial intelligence (AI) to connect the data through machine learning (ML) algorithms and deep neural networks (DNNs) [11]. While DFT-based calculations have provided important input data for CALPHAD modeling [12], it is currently still necessary to refine the CALPHAD model parameters using experimental data in order to accurately reproduce experimental observations, particularly phase transitions [13,14]. The need of such refinements significantly hinders the computational discovery and design of materials. To fully understand the differences between DFT-based calculations and CALPHAD modeling, it is necessary to dive deep into their fundamentals and build connections so that in the future the CALPHAD model parameters can be evaluated solely from the DFT-based calculations with experiments as the validation of predictions.

Thermodynamics is a science concerning the state of a system, whether it is stable, metastable, or unstable, when interacting with its surroundings. The interactions can involve exchanges of any combinations of heat, work, and mass between the system and the surroundings, defined by the boundary conditions. The typical work includes contributions from the external mechanic, electric and magnetic fields. Thermodynamics is commonly divided into four branches, i.e., classical Gibbs thermodynamics, Statistical thermodynamics, quantum thermodynamics, and irreversible thermodynamics. In a recent overview article [15], the author discussed fundamental thermodynamics, thermodynamic modeling, and the applications of computational thermodynamics. In another recent perspective article [16], the author focused on irreversible thermodynamics as part of a more comprehensive framework of thermodynamics.

In the present paper, the fundamentals of thermodynamics will be reviewed through the derivation of the combined law of thermodynamics with the entropy production due to internal processes and thus without the constraint of equilibrium. The four branches of thermodynamics will then be discussed individually along with their contributions to the combined law of thermodynamics and their integration into a holistic view of thermodynamics. At the end, the author's perspectives on the CALPHAD modeling in the next 50 years will be discussed.

## 2 Review of the fundamentals of thermodynamics

The fundamentals of thermodynamics are centered on the first and second laws of thermodynamics and their combination into the combined law of thermodynamics. Since the first and second laws of thermodynamics are represented by an equality and an inequality, respectively, they had remained separately until Gibbs combined them to create the combined law of thermodynamics [17–19]. The first law of thermodynamics describes the interactions between the system and its surroundings and stipulates that the exchange of energy between the system and its surroundings is balanced by the internal energy change of the system. While the second law of thermodynamics governs the internal processes inside the system under those interactions and states that any spontaneous internal processes are irreversible and must produce entropy. In the 1870's Gibbs [17] combined them together to create the combined law of thermodynamics and called it the fundamental thermodynamic equation [19].

However, Gibbs considered only the case when the inequality is replaced by an equality, i.e., when the second law of thermodynamics vanishes for a system and was thus effectively removed from the Gibbs combined law of thermodynamics. Furthermore, Gibbs combined law of thermodynamics was first derived for a closed system without mass exchange with the surroundings, which was added later into the combined law of thermodynamics by introducing the chemical potential abruptly for each independent component of the system. This has caused considerable confusion in the literature on the concept of the chemical potential.

In this section of the present paper, these two issues are addressed in deriving the combined law of thermodynamics of an open system with internal processes, and more detailed discussions can be found in these books [20,21]. It is noted that Gibbs combined law of thermodynamics with the internal energy of the system as a function of entropy and volume inspired Maxwell to construct manually a three-dimensional model to represent the internal energy surface as a function of entropy and volume with one copy sent to Gibbs [19] and one kept in Cavendish Laboratory at the University of Cambridge as shown in Figure 1.

*Figure 1: Photo of the three-dimensional model in Cavendish Laboratory at the University of Cambridge made by Maxwell to represent the internal energy surface as a function of entropy and volume.*

## 2.1 First law of thermodynamics

To properly introduce chemical potential in the combined law of thermodynamics, let us consider a system that is free to exchange heat, work, and mass with the surrounding and write the first law of thermodynamics as follows

$$dU = dQ + dW + \sum_{i=1}^{c} U_i dN_i \qquad \textit{Eq. 1}$$

where $dU$ is the internal energy change of the system, $dQ$, $dW$, and $dN_i$ are the exchanges of heat, work, and moles of component $i$ from the surroundings to the system, with the work including mechanical, electric, and magnetic work, $c$ is the number of independent components, and $U_i$ is the partial internal energy of component $i$ defined as follows,

$$U_i = \left(\frac{\partial U}{\partial N_i}\right)_{dQ=0, dW=0, N_{j \neq i}} \qquad \textit{Eq. 2}$$

It is noted that the first law of thermodynamics does not prescribe whether the system is in internal equilibrium or not, i.e., independent of what happens inside the system. Consequently, the value of $U_i$ in the system may be different from that in the surroundings, while the exchanges of heat and work are independent of the state of the system. As shown below, the inclusion of mass exchange in the first law of thermodynamics enables the nature introduction of chemical potential.

## 2.2 Second law of thermodynamics

Gibbs [17] followed Clausius' definition of entropy exchange ($d_Q S$) between an equilibrium system and its surrounding with only reversible heat exchange as follows

$$d_Q S = \frac{dQ}{T} \qquad \textit{Eq. 3}$$

For a nonequilibrium system, the second law of thermodynamics stipulates that an irreversible internal process ($ip$) inside the system generates positive entropy production, $d_{ip}S$, as follows

$$d_{ip}S \geq 0 \qquad \textit{Eq. 4}$$

where the equality represents that there are no internal processes in the system, indicating that the system is either at equilibrium or under a freezing-in condition as discussed by Hillert [20] and to be further discussed below.

For an open system with mass exchange between the system and its surroundings, each mass exchange carries entropy exchange at the same time. Consequently, the total entropy change of the system can be written as follows [15,16,20,21]

$$dS = d_Q S + \sum\nolimits_{i=1}^{c} S_i dN_i + d_{ip}S = \frac{dQ}{T} + \sum\nolimits_{i=1}^{c} S_i dN_i + d_{ip}S \qquad \textit{Eq. 5}$$

where $S_i$ is the partial entropy of component *i* defined as

$$S_i = \left(\frac{\partial S}{\partial N_i}\right)_{dQ=0, d_{ip}S=0, N_{j\neq i}} \qquad \textit{Eq. 6}$$

The work exchange between the system and its surroundings does not enter Eq. 5 directly, but indirectly affects the entropy change of the system by introducing internal processes. It is important to note that the first two terms in Eq. 5 concern the exchanges between the system and its surroundings, while the third term does not, so that the total entropy change contains the contributions from both internal processes and exchanges between the system and its surroundings. Therefore, $dS$ can be either positive or negative, which is not in contradiction with the second law of thermodynamics as the second law of thermodynamics concerns only the entropy production of an independent internal process represented by Eq. 4 or the last term in Eq. 5.

### 2.3 Combined law of thermodynamics

Consequently, the combined law of thermodynamics form with internal processes can be obtained by combining Eq. 1 and Eq. 5 as follows [15,16,20,21]

$$dU = TdS + dW + \sum_{i=1}^{c} \mu_i dN_i - Td_{ip}S = \sum_{a=1}^{n} Y^a dX^a - Td_{ip}S \qquad Eq.\ 7$$

$$\mu_i = U_i - TS_i = \left(\frac{\partial U}{\partial N_i}\right)_{d_{ip}S=0, X^a \neq N_i} \qquad Eq.\ 8$$

where $\mu_i$ is the chemical potential of component $i$, $n$ is the total number of independent contributions to the internal energy of the system controlled from the surroundings, i.e., the number of external variables, and $Y^a$ and $X^a$ represent the pairs of conjugate variables with $Y^a$ for potentials, such as temperature, stress or pressure, electrical and magnetic fields, and chemical potential, and $X^a$ for molar quantities, such as entropy, strain or volume, electrical and magnetic displacements, and moles of components. The concept of the chemical potential is thus naturally introduced by Eq. 8, containing the contributions from both partial internal energy and partial entropy of the component. The use of superscript in $Y^a$ and $X^a$ is to indicate that they are related to the exchange between the system and its surroundings, thus external variables, though entropy contains both internal and external contributions as shown by Eq. 5. The constraint of constant entropy production in Eq. 8 with heat exchange between the system and its surroundings signifies its difference from Eq. 2 which is for an adiabatic system even though both are for the change of internal energy with respect to the change of the amount of the component $i$.

In a system, there can be many independent internal processes. Each of them must result in a positive entropy production, which can be divided into four parts: (1) heat generation $(d_{ip}Q)$, (2) consumption of some components as reactants $(dN_{r,j})$, (3) production of some components as products $(dN_{p,k})$, and (4) reorganization of its configurations $(d_{ip}S^{config})$, as follows [16,22],

$$d_{ip}S = \frac{d_{ip}Q}{T} - \sum_{j} S_j dN_{r,j} + \sum_{k} S_k dN_{p,k} + d_{ip}S^{config} = \frac{D}{T} d\xi \qquad \textit{Eq. 9}$$

where $S_j$ and $S_k$ are the partial entropies of reactant $j$ and product $k$ of the internal process. The last part of Eq. 9 is based on the *linear proportionality* approximation with $D$ and $d\xi$ being the driving force and progress of the internal process that is represented by a molar quantity or a combination of molar quantities involved in the internal process. In reality, internal processes do not obey linear proportionality, but in principle, one can always select a small enough $d\xi$ so the higher-order terms are less important and then perform integrations along the pathways to obtain overall entropy production. On the other hand, to study the stability of an internal process or a system such as the limit of instability and critical points, one would need to include higher order terms beyond linear proportionality. The second law of thermodynamics prescribes that any spontaneous process with $d\xi > 0$ must have a positive driving force, i.e., $D > 0$, to result in a positive entropy production. The internal energy is thus a function of all $X^a$ and internal variables of $\xi$, i.e., $U(X^a, \xi)$, so are all the potentials, i.e., $Y^a(X^b, \xi)$ where $X^b$ denotes all molar quantities including $X^a$. It is noted that this is not in contradiction to the early statement that the first law of thermodynamics does not concern the state of the systems, i.e., $\xi$, and the dependence of internal energy on $\xi$ is introduced due to the dependence of entropy on $\xi$ for a non-equilibrium system.

For equilibrium systems, Eq. 5 and Eq. 7 are reduced to the following equations

$$dS = \frac{dQ}{T} + \sum_{i=1}^{c} S_i dN_i \qquad Eq.\ 10$$

$$dU = \sum_{a=1}^{n} Y^a dX^a \qquad Eq.\ 11$$

The internal energy is thus a function of all $X^a$ only, i.e., $U(X^a)$, and the internal variables of $\xi$ become dependent variables as functions of $X^a$ obtained from the condition of $Dd\xi = 0$. It is noted that at equilibrium with $D = 0$, the external variables determines all the internal variables, i.e., $\xi$ becomes a dependent variable. For systems under a freezing-in condition with $d\xi = 0$ and $D > 0$, $\xi$ is an independent variable.

## 3 Classical Gibbs thermodynamics

### 3.1 Combined law of thermodynamics for equilibrium systems

Gibbs [17,18] first considered closed equilibrium systems under hydrostatic pressure ($P$), i.e., $dN_i = 0$ and $d_{ip}S = 0$, and Eq. 10 and Eq. 11 are thus reduced to the following equations as shown in all textbooks

$$dS = \frac{dQ}{T} \qquad Eq.\ 12$$

$$dU = TdS - PdV \qquad Eq.\ 13$$

where the negative sign in front of pressure is because $-PdV$ is the work added to the system with $P$ and $V$ changing in opposite directions in a stable system as discussed below. The internal energy is thus a function of entropy and volume, i.e., $U(S, V)$, which is the model that Maxwell constructed as shown in Figure 1.

For open systems, Gibbs directly added the chemical potential term to Eq. 13 as follows

$$dU = TdS - PdV + \sum_{i=1}^{c} \mu_i dN_i \qquad \textit{Eq. 14}$$

The internal energy is thus a function of entropy, volume, and moles of each component, i.e., $U(S, V, N_i)$, which are all molar quantities and termed as natural variables of $U$. By virtual internal processes of moving infinitesimal amount of $S$, or $V$, or $N_i$ at various locations inside the system, it can be shown that the equilibrium is reached when the driving force for each internal process is zero [21]. As the driving force is denoted by the difference of the conjugate potential of the internal process, every potential, i.e., $T$, $P$, and $\mu_i$, which are the first partial derivatives of $U$ with respect to its natural variables, is homogeneous everywhere in the system at equilibrium. The derivatives above and throughout the paper are all partial derivatived and performed with other natural variables kept constant unless specified otherwise. The stability of the equilibrium is further dictated by the second derivatives of $U$, i.e., the first derivative between conjugate potentials and molar quantities, as follows

$$\frac{\partial^2 U}{\partial \xi^2} = \frac{\partial^2 U}{\partial (X^a)^2} = \frac{\partial Y^a}{\partial X^a} > 0 \qquad \textit{Eq. 15}$$

While the internal processes with $d\xi = dS$ and $d\xi = dV$ can be independently carried out, the internal process with $d\xi = dN_i$ cannot be performed independently because it will carry the changes of entropy and volume simultaneously as follows

$$dS = S_i dN_i \qquad Eq.\ 16$$

$$dV = V_i dN_i = \frac{\partial V}{\partial N_i} dN_i \qquad Eq.\ 17$$

where $S_i$ is defined by Eq. 6, and $V_i$ is the partial volume of component $i$. These will induce two internal processes in opposite directions if the homogenous temperature and pressure are to be maintained in the system under equilibrium. This complication can be simplified by defining the Gibbs energy and re-writing the combined law of thermodynamics as follows

$$dG = d(U - TS + PV) = -SdT + VdP + \sum_{i=1}^{c} \mu_i dN_i \qquad Eq.\ 18$$

Gibbs energy is thus with $T$, $P$ and $N_i$ as its natural variables with two being potentials, i.e., $G(T, P, N_i)$.

### 3.2 Gibbs-Duhem equation

From Eq. 14, it is easy to show the following equation for a homogeneous equilibrium system

$$U = TS - PV + \sum_{i=1}^{c} \mu_i N_i \qquad Eq.\ 19$$

By moving $TS$ to the left side of Eq. 19, one defines the Helmholtz energy with the corresponding combined law of thermodynamics shown below

$$dF = d(U + PV) = -SdT - PdV + \sum_{i=1}^{c} \mu_i dN_i \qquad Eq.\ 20$$

Helmholtz energy is thus with $T$, $V$ and $N_i$ as its natural variables with one being potential, i.e., $F(T, V, N_i)$. By moving $TS$ and $-PV$ to the left side, one obtains the Gibbs energy as shown by Eq. 18. It is noted that "free" is not used here for Gibbs energy and Helmholtz energy as recommended by International Union of Pure and Applied Chemistry (IUPAC) [23,24], while in the present work, the free energy is used to denote all energies derived from the internal energy.

If all terms in the right-hand side of Eq. 19 are moved from right to left, one obtains

$$\Phi = U - \left(TS - PV + \sum_{i=1}^{c} \mu_i N_i\right) = 0 \qquad \textit{Eq. 21}$$

The differentiation of Eq. 21 together with Eq. 14 gives the Gibbs-Duhem equation

$$d\Phi = -SdT + VdP - \sum_{i=1}^{c} N_i d\mu_i = 0 \qquad \textit{Eq. 22}$$

The natural variables of $\Phi$ are thus all potentials, i.e., $\Phi(T, P, \mu_i) = 0$. It is tempting to call this free energy as the *Duhem energy* even though it is zero, which can be written in a more general form as

$$d\Phi = d\left(U - \sum_{a=1}^{n} Y^a dX^a\right) = -\sum_{a=1}^{n} X^a dY^a = 0 \qquad \textit{Eq. 23}$$

It also needs to emphasize that the Gibbs-Duhem equation is only applicable to a homogeneous equilibrium system or a portion of the equilibrium system, such as a homogeneous phase discussed in the next section.

### 3.3 Gibbs phase rule

The significance of Eq. 22 or Eq. 23 is that the potentials in a homogeneous equilibrium system are not independent of each other. In a system with $n$ pairs of conjugate variables, i.e., $X^a$ and $Y^a$, there are $n$ independent variables. They can be all molar quantities such as the internal energy shown by Eq. 11 or some combinations of potentials and molar quantities such as Gibbs energy (Eq. 18) with two potentials and Helmholtz energy (Eq. 20) with one potential. If $n$ independent variables are all potentials, one obtains the Gibbs-Duhem equation (Eq. 23) or *Duhem* energy with the value being zero. This indicates that at least one independent variable of an equilibrium system must be a molar quantity.

For a heterogeneous system with two or more homogeneous phases in equilibrium with each other, Gibbs [17,18] discussed the geometry of phase relations with the axes being molar quantities. Since phase equilibria are defined by homogeneous potentials in the system, it is easier to understand phase relations if potentials are used as axis variables. Let us consider a homogeneous phase, say $\beta$, and apply the Gibbs-Duhem equation to it as follows

$$d\Phi = -\sum_{a=1}^{n} X_a^{\beta} dY_a = 0 \qquad \textit{Eq. 24}$$

where the subscripts represent the properies of a phase inside the system.

For a system with $p$ phases co-existing in equilibrium, Eq. 24 needs to be applied to each phase, resulting in $p$ equations of Eq. 24, noting that the molar quantities are different in individual phases. The number of potentials that can be varied independently without changing the number of phases in equilibrium is thus

$$\upsilon = n - p \qquad \textit{Eq. 25}$$

The maximum number of phases that can co-exist in equilibrium with $\upsilon$=0 is then

$$p_{max} = n \qquad \textit{Eq. 26}$$

It should be emphasized that in many textbooks, $\upsilon$ is termed as degrees of freedom or independent variables of the system. This is inaccurate or at least not rigorous because the number of independent variables in the system is always $n$ as defined by the combined law, while $\upsilon$ defines the number of independent potentials without changing the number of phases in equilibrium. For a system with $p = p_{max} = n$ thus $\upsilon = 0$, all $n$ $X^a$ can still be changed independently within certain ranges that will change the relative amounts of each phase but not the number of phases co-existing in equilibrium.

### 3.4 Phase diagrams

As discussed above, every potential has the same value in all phases co-existing in equilibrium, and phase diagrams with potentials as axes are thus important. Eq. 24 depicts that each phase is represented by a *(n-1)*-dimensional feature in the space of *n* potentials. A two-phase equilibrium is thus the *(n-2)*-dimensional feature where two *(n-1)*-dimensional features intersect each other. Following the Gibbs phase rule, a $(p_{max} = n)$-phase equilibrium is 0-dimension point where *n* *(n-1)*-dimensional features intersect each other, commonly referred as invariant point because the values of all potentials are fixed. It is thus evident that Gibbs phase rule can be directly applied to potential phase diagrams.

To display more information about phases, it is useful to change one or more potentials to their conjugate molar quantities. Since a molar quantity has different values in different phases, the dimensionality of each phase region discussed above increases by one when one potential is replaced by its conjugated molar quantity until the dimension reaches *n*. The phases in equilibrium are then connected by tie-lines with their ends represent the values of their respective molar quantities. When all axes of the phase diagram are converted to molar quantities, the dimension of every phase region is the same and equals to the number of axes, i.e., *n*, and any features with lower dimensionality are termed as phase boundaries. The molar properties of individual phases are connected through the lever rule as follows

$$X_a^0 = \sum_{\beta=1}^{p} f^\beta X_a^\beta \qquad \textit{Eq. 27}$$

where $X_a^0$ and $X_a^\beta$ are the values of the over-all $X_a$ in the system and in the $\beta$ phase where the subscript is used for quantities of a phase inside the system similar to Eq. 24, respectively, and $f^\beta$ the mole fraction of the phase $\beta$ with summation going over all phases in equilibrium in the system.

One has to section a multidimensional phase diagram in order to visualize it in two dimensions. Sectioning a potential diagram decreases the total number of potentials and does not change the features of the resulting phase diagram. The Gibbs phase rule thus becomes

$$\beta = n - p - n_s \qquad \textit{Eq. 28}$$

$$p_{max} = n - n_s \qquad \textit{Eq. 29}$$

where $n_s$ denotes how many times the potential diagram is sectioned, i.e., the number of potentials being fixed. However, sectioning a phase diagram with molar quantities does not change $p_{max}$ and $\upsilon$ and is thus more complicated because tie lines are usually not in the resulting phase diagram, which is discussed in detail by Hillert [20].

Commonly used phase diagrams are the $T - P$ potential phase diagram for one component systems, $T - x_B$ under constant $P$ for binary systems with $x_B$ being the mole fraction of component $B$, $x_B - x_C$ section under constant $T$ and $P$ for ternary systems with $x_B$ and $x_C$ being the mole fractions of component $B$ and $C$ (termed as isothermal sections), and $T - x_i$ section of ternary or multi-component systems under constant $P$ and compositions of other components (termed as isopleth). It is noted that the lever rule, i.e., Eq. 27, cannot be directly used in an isopleth as the tie-lines are typically not in the phase diagram. When tie lines happen to be in the isopleth, the isopleth is often called pseudo-lower-order phase diagram such as pseudo-binary or pseudo-ternary phase diagrams as they behavior like binary or ternary systems.

## 4 Gibbs and quantum statistical thermodynamics

Statistical mechanics was introduced by Gibbs in 1901 [25] based on the foundations established by Clausius, Boltzmann, and Maxwell. Gibbs considered "a great number of independent systems (states) of the same nature (of a system), but differing in the configurations and velocities which they have at a given instant, and differing not merely infinitesimally, but it may be so as to embrace every conceivable combination of configuration and velocities"[25]. He thus broadened the early statistical mechanics from the consideration of the particles of a system to independent systems (configurations of a system). Gibbs systematically discussed the fundamental equation of statistical mechanics in terms of the principle of conservation of probability, applied it to the theory of errors in the calculated state of a system and the integration of the differential equations of motion, and studied the system under statistical equilibrium with ensembles in which the logarithm of the probability of state is a linear function of the energy. A differential equation relating to average values in the ensemble was found to be identical in form with the fundamental differential equation of thermodynamics with the average index of probability of state corresponding to the entropy with change of sign and the modulus to temperature. By using the combined law of thermodynamics in terms of the internal energy [19] and differentiating the internal variables in the system and the external variables from the surroundings, Gibbs [25] considered the internal variable entropy due to the distribution of various microstates in the system but not the entropy of each microstate itself. Gibbs was then able to define the probability and the now-named partition function of each configuration in the system.

In formulating statistical mechanics in the framework of quantum mechanics, Landau and Lifshitz [26] considered a closed system in complete statistical equilibrium by dividing it into a large number of macroscopic subsystems. They emphasized that the entropy of a closed system in complete statistical equilibrium can also be defined directly, without dividing the system into subsystems by imagining that the system considered is actually only a small part of a fictitious very large system. Landau and Lifshitz [26] introduced the number of quantum states corresponding to the energy interval equal in the order of magnitude to the mean fluctuation of energy of the system and showed the entropy of the system in terms of the tracer of each quantum state. By correlating the number of quantum states with the particle states in the limit of the classical theory, they obtained the entropy of a system as follows

$$S = -k_B \sum_{k=1}^{m} p^k lnp^k \qquad \textit{Eq. 30}$$

where $m$ is the number of configurations, the probability $p^k$ of configuration $k$ was denoted by $w_n$ in their Eq. 7.10, and $k_B$ is added here to be consistent with the current convention.

Landau and Lifshitz [26] thus obtained the Gibbs distribution and presented the partition function of the system, $Z$, in relation to the Helmholtz energy of the system, $F$, and the internal energy of each quantum state, $E^k$, in their Eqs. 31.1 to 31.4, which are re-written as follows

$$Z = e^{-\frac{F}{k_BT}} = \sum_{k=1}^{m} e^{-\frac{E^k}{k_BT}} = \sum_{k=1}^{m} Z^k \qquad \textit{Eq. 31}$$

$$F = -k_BTlnZ + k_BT\left(\sum_{k=1}^{m} p^k lnZ^k - \sum_{k=1}^{m} p^k lnZ^k\right)$$
$$= \sum_{k=1}^{m} p^k E^k + k_BT\sum_{k=1}^{m} p^k lnp^k = \sum_{k=1}^{m} p^k E^k - TS \qquad \textit{Eq. 32}$$

$$p^k = \frac{Z^k}{Z} = \frac{e^{-\frac{E^k}{k_BT}}}{Z} = e^{-\frac{E^k-F}{k_BT}} \qquad \textit{Eq. 33}$$

It is important to emphasize that Landau and Lifshitz [26] used quantum states in the above equations, while Gibbs did not do so as quantum mechanics was not developed at that time. Let us consider a hypothetical system with only one configuration, Eq. 32 thus becomes

$$F = E^k \qquad \textit{Eq. 34}$$

Since $F = E^k - TS^k$ by definition, Eq. 34 gives $S^k = 0$ at finite temperature, indicating the configurations in Eq. 30 to Eq. 33 are all pure quantum states with only one configuration each as envisioned by Landau and Lifshitz [26]. For systems of practical interest, the number of pure quantum states is very large, and their complete sampling is in general intractable. The current available solution is their coarse graining through DFT [5,6] as discussed below, resulting in a non-zero entropy for each configuration at finite temperature and the necessity to modify the formula of the partition function as discussed in Section 7 in terms of zentropy theory.

It is noted that the statistical equilibrium of a closed system is usually discussed in terms of thermal equilibrium between the system and a thermal bath (surroundings). As pointed out recently by the author [16], thermal fluctuation in a closed statistically equilibrated system is an internal process and results in the heat or work exchange between the system and its surroundings. The entropy production of thermal fluctuation can be represented by Eq. 9 as follows

$$d_{ip}S = \frac{d_{ip}Q}{T} + d_{ip}S^{config} \geq 0 \qquad \textit{Eq. 35}$$

The thermal fluctuation either releases heat ($d_{ip}Q > 0$) and makes the system more ordered ($d_{ip}S^{config} < 0$) or absorbs heat ($d_{ip}Q < 0$) and makes the system more disordered ($d_{ip}S^{config} > 0$) with the corresponding amount of heat exchange $\left(dQ = -d_{ip}Q\right)$ between the system and its surroundings to maintain $d_{ip}S = 0$. For an isolated system without a thermal bath, $dQ = 0$, the system reaches the internal statistical equilibrium with the equality in Eq. 35, i.e., all internal processes are reversible with the following relation between the heat production and the change of internal configurations of each thermal fluctuation to give $d_{ip}S = 0$,

$$d_{ip}Q = -Td_{ip}S^{config} \qquad \textit{Eq. 36}$$

One example of thermal fluctuation is the Brownian motion. The internal process releases hear with $d_{ip}Q > 0$ when the atoms return to their equilibrium positions, while absorbes heat with $d_{ip}Q < 0$ is when the atoms fluctuate away from their equilibrium positions. In the former case, the released heat can be extracted through an external effort to perform certain amount of work, which unfortunately has been considered in the literature as the evidence of the microscopic 'violation of second law of thermodynamics'. The misunderstanding is because the $d_{ip}S^{config}$ was missed in counting the entropy changes of the internal processes as discussed in detail by the author recently [16].

## 5 Quantum thermodynamics in the framework of density functional theory

Quantum mechanics provides a description of the physical properties of nature at the scale of atoms and subatomic particles. A solid can be thought of as a collection of interacting positively charged nuclei and negatively charged electrons and can theoretically be treated by solving the many-body Schrödinger quantum mechanics equation involving both the nuclei and the electrons [4]. However, it is extremely difficult to solve the equation due to its many-body nature with too many electrons. On the other hand, DFT developed in 1960's [5,6] aims to represent the outcome of those interactions by a single wave function of one electron and articulates that there is a ground state in each system at zero K defined by a unique electron density. The widely used DFT-based calculations represents the state-of-the-art solution of the multi-body Schrödinger equation with several approximations as follows [12].

- Adiabatic or Born-Oppenheimer approximation [27]. The nuclei that are much heavier than the electrons are assumed to be "frozen" and only contribute to an external potential for the electrons. The electrons are always in an instantaneous ground state with the nuclei.
- Independent-electron approximation. Each electron moves independently of the others in an average effective potential collectively determined by the nuclei and all electrons. DFT by Hohenberg and Kohn [5] is formulated as an exact theory of many-body systems. It articulates that for an interacting electron gas there exists a universal functional of the density such that the energy is at its minimum value, i.e., the ground-state energy with a unique ground-state electron density. The Kohn-Sham approach [6] explicitly separates the independent-electron kinetic energy and long-range Coulomb interaction energy and replaces the many-body electron problem using independent electrons with an exchange-correlation functional of the electron density and an associated exchange-correlation energy and potential, i.e., *coarse graining of electrons*. Consequently, the exchange-correlation energy can be approximated as a local functional of the electron density.

- Exchange-correlation functional approximation by the local spin density approximation (LSDA) [28,29] and the generalized gradient approximation (GGA) [30–32]. In LSDA, the exchange-correlation energy density at each point in space is assumed to be the same as in a homogenous electron gas with the same electron density. While in GGA, the exchange-correlation energy density depends additionally on the gradient of the electron density.
- Replacement of the strong Coulomb potential of the nucleus and the tightly bound core electrons by a pseudopotential. This pseudopotential represents an effective potential acting on valence electrons and is obtained from atomic calculations. Since it is not unique, it is be tailored to simplify calculations such as the commonly used ultrasoft pseudopotentials and the projector augmented wave (PAW) method [33].

With above approximations, the DFT-based first-principles calculations solve a set of one-electron Schrödinger's equations, one for each valence electron in the system with supercells of atomic structures and periodic boundary conditions, to obtain the ground-state electron density, which is used to obtain ground-state energy and other properties of the system at 0 K. Additional calculations at volumes around that of the ground-state configuration can be performed to obtain the equation of states (EOS) along with the bulk modulus and its derivative with respect to volume by fitting the energy as a function of volume using various EOS models [34].

As the third law of thermodynamics postulates, the entropy of a system equals zero at 0 K, so is the entropy of the ground-state configuration at 0 K. At finite temperature, electronic structures will change, and nuclei will vibrate, resulting in the increase of entropy. Kohn and Sham [6] used the finite temperature generalization of ground-state energy of an interacting inhomogeneous electron gas by Mermin [35] and formulated the entropy of thermal electrons at finite temperature. Wang et al. [36] added the vibrational contribution and presented the free energy as follows

$$F^k = E^{k,0} + F^{k,el} + F^{k,vib} = E^k - TS^k \qquad \textit{Eq. 37}$$

$$E^k = E^{k,0} + E^{k,el} + E^{k,vib} \qquad \textit{Eq. 38}$$

$$S^k = S^{k,el} + S^{k,vib} \qquad \textit{Eq. 39}$$

where $F^k$, $E^k$, and $S^k$ are the Helmholtz energy, internal energy, and entropy of configuration $k$, $F^{k,el}$, $E^{k,el}$, and $S^{k,el}$ are the contributions of thermal electron to Helmholtz energy, internal energy, and entropy of configuration $k$ based on the Fermi–Dirac statistics for electrons, and $F^{k,vib}$, $E^{k,vib}$, and $S^{k,vib}$ are the vibrational contributions to Helmholtz energy, internal energy, and entropy of configuration $k$ based on the Bose–Einstein statistics for phonons, respectively. The vibrational contributions can be obtained by either phonon calculations or Debye model with the former being more accurate and the latter more efficient [36,37] through the high throughput DFT Tool Kits (DFTTK) [38,39]. The detailed equations for these quantities can be found in the literature [21,36]. For vibration-induced dipole-dipole long-range interactions, we have developed a mixed-space approach with the short-range interactions accounted for by supercells in the real space, the analytical solution for the origin in the reciprocal space which represents the infinite in the real space, and an interpolation scheme between them [40–42].

Non-ground-state configurations can be created by systematically varying the internal degrees of freedom of the ground-state configuration of the system. For stable non-ground-state configurations, Eq. 37 to Eq. 39 can be used to predict their Helmholtz energies. For unstable non-ground-state configurations, there will be imaginary frequencies in their phonon dispersion curves, and their Helmholtz energies can thus not be directly predicted by means of phonon calculations. Even though the Debye model does not explicitly concern whether the configuration is stable or unstable, the physical significance of those predictions needs to be further investigated [43]. This is the central topic in integrating the DFT-based calculations into the CALPHAD modeling [15,44–52], which will be further discussed in Sections 8.

There have been continued efforts in improving DFT methods to obtain better electron density and total energy [53–60] include time-dependent DFT (TDDFT) [61–63], random phase approximation (RPA) [64,65], density-matrix functional theory (DMFT) [66–69], DFT+U [70–72], dynamical mean-field theory [73,74], benchmarking with experimental measurements [75], deep neural network machine learning models [76–78], and some other hybrid methods [79]. These approaches primarily aim to improve the calculations for the ground-state configuration through approximations for the exchange correlation functional, i.e., Eq. 37 to Eq. 39. However, they may not be able to capture the statistical contributions from non-ground-state configurations reflected by experimental observations performed at finite temperatures as shown by Eq. 31 to Eq. 33. This will be further discussed in Section 7.

It is noted that recently Perdew and co-workers pointed out that symmetry breaking can arise when a wave-like fluctuation drops to zero frequency [80]. By considering approximate density functional for exchange and correlation that breaks symmetry, they demonstrated that symmetry breaking with an advanced density functional might reliably describe strong correlation and can thus be more revealing than an exact functional that does not [80,81]. They discussed that the ground-state total energy of a system of interacting electrons has contributions from fluctuations of various wavevectors, including the nonnegative zero-point energies of its collective excitations. For the uniform electron gas, they developed a new approximate density functional that includes the time-dependent fluctuations using the fluctuation-dissipation theorem as discussed in detail in ref. [80], which is currently not practical for real systems due to the need to sum over all unoccupied and occupied orbitals to find the non-interacting linear response function and the lack of reliable exchange-correlation kernel. It is noted that the ground-state and non-ground-state configurations discussed above do have different symmetries.

# 6 Irreversible thermodynamics in terms of internal processes

## 6.1 Classical and extended irreversible thermodynamics

Typical approaches to irreversible thermodynamics in the literature start from the Gibbs thermodynamics, i.e., Eq. 13 or Eq. 14 for closed systems with hydrostatic pressure or Eq. 11 in general. For example, Onsager used the microscopic reversibility that requires that if $\alpha$ and $\beta$ be two quantities which depend only on the configuration of molecules and atoms, the event $\alpha = \alpha'$ followed some seconds later by $\beta = \beta'$ will occur as often as the event $\beta = \beta'$, followed later by $\alpha = \alpha'$. It also requires the same if $\alpha$ and $\beta$ depend on the velocities of elementary particles in such a manner that they are not changed when the velocities are reversed. Onsager noted that "the principle of microscopic reversibility is less general than the fundamental laws of thermodynamics", which is shown below to be incompatible with irreversible processes. Based on the principle of microscopic reversibility, Onsager further derived the reciprocal relation among kinetic coefficients of fluxes which will be discussed in Section 6.3.

While more often, as shown by de Groot and Mazur[82], Kondepudi and Prigogine [83], and in most textbooks, the combined law by Gibbs with $d_{ip}S = 0$ is used directly to derive formulas for irreversible thermodynamics with Eq. 14 re-written as, neglecting convection and volumetric flow, [83]

$$dS = \frac{1}{T}dU - \sum_{i=1}^{c} \frac{\mu_i}{T} dN_i = d_e S \qquad \textit{Eq. 40}$$

where the last portion was added in the present work with $d_e S$ being the entropy change or entropy current between the system and its surroundings as used by Kondepudi and Prigogine [83] who also used the symbol $\boldsymbol{J_S}$, or between an internal process and its surrounding inside the system. Its time-dependent form can be written as

$$\frac{dS}{dt} = \dot{S} = \frac{1}{T}\dot{U} - \sum_{i=1}^{c} \frac{\mu_i}{T} \dot{N}_i = {}_e\dot{S} \qquad \textit{Eq. 41}$$

The time-dependent first law of thermodynamics was then formulated in a flux form, re-arranged, and inserted back to Eq. 41,

$$\dot{U} = \dot{Q} + \sum_{i=1}^{c} U_i \dot{N}_i = -\nabla \bullet \left(\boldsymbol{J_Q} + \sum_{i=1}^{c} U_i \boldsymbol{J_i}\right) = -\nabla \bullet (\boldsymbol{J_U}) \qquad \textit{Eq. 42}$$

$$\dot{S} = {}_e\dot{S} = -\nabla \bullet \left(\frac{\boldsymbol{J_U}}{T} - \sum_{i=1}^{c} \frac{\mu_i \boldsymbol{J_i}}{T}\right) + \boldsymbol{J_U} \bullet \nabla\left(\frac{1}{T}\right) - \sum_{i=1}^{c} \boldsymbol{J_i} \bullet \nabla\left(\frac{\mu_i}{T}\right) \qquad \textit{Eq. 43}$$

where $\boldsymbol{J_Q}$ and $\boldsymbol{J_i}$ are the fluxes of heat and component $i$, respectively. This is a circular operation, resulting in the second law of thermodynamics with the first law of thermodynamics effectively removed from the combined law, i.e., Eq. 10, as follows

$$\dot{S} = \frac{1}{T}\dot{Q} + \sum_{i=1}^{c} S_i \dot{N}_i = {}_e\dot{S} \qquad \textit{Eq. 44}$$

This in principle should not result in any new information related to irreversible processes or the second law of thermodynamics as discussed below.

It should be emphasized that the first law of thermodynamics represented by Eq. 42 is applicable to both equilibrium and non-equilibrium systems with internal processes when it was introduced in Eq. 1, while Eq. 40 and Eq. 41 are for equilibrium systems only without internal processes, so Eq. 43 is only applicable to equilibrium systems. To solve this problem of circular operation, the common next step in the literature is to separate Eq. 43 into two contributions with the first term for entropy change or entropy current and the combination of the second and third terms as the entropy production due to internal processes i.e.,

$$\dot{S} = {}_e\dot{S} = {}_e\dot{S} + {}_{ip}\dot{S} \qquad \textit{Eq. 45}$$

where ${}_{ip}\dot{S}$ is the entropy production rate due to irreversible internal processes, written as ${}_i\dot{S}$ or $\sigma$ by Kondepudi and Prigogine [83]. This is evidently incorrect because one cannot start with the equation applicable to equilibrium systems without internal processes and end up with an equation for non-equilibrium systems with internal processes because Eq. 45 implies that ${}_{ip}\dot{S} = 0$.

The problem is due to the circular use of the first law of thermodynamics and the artificial separation of contributions because the internal energy change does not concern the internal processes, but only the exchange between the system and its surroundings as defined by the first law of thermodynamics. This is shown by Eq. 7 in which $Td_{ip}S$ is added into $TdS$ in Eq. 5, and it should thus be emphasized that the internal processes do not change the internal energy of the system. The second law of thermodynamics concerns exclusively the entropy production due to irreversible internal processes.

Another development in thermodynamics is the extended irreversible thermodynamics [84,85], which aims to describe phenomena at frequencies comparable to the inverse of the relaxation times of the fluxes by including the fast variables among the set of basic independent variables. This is in analogy to keep the entropy production term as shown in Eq. 7, which can be re-arranged as follows

$$\dot{S} = \frac{1}{T}\dot{U} - \frac{1}{T}\dot{W} - \sum_{i=1}^{c} \frac{\mu_i}{T}\dot{N}_i + {}_{ip}\dot{S} = {}_{e}\dot{S} + {}_{ip}\dot{S} \qquad \textit{Eq. 46}$$

The central question is then how to formulate $d_{ip}S$ or ${}_{ip}\dot{S}$. In the extended irreversible thermodynamics, $d_{ip}S$ is formulated as a function of fluxes, and ${}_{ip}\dot{S}$ is then related to the divergency of fluxes [84,85]. Since the unit of $d_{ip}S$ does not contain time, it seems awkward to include fluxes as its independent variable. As shown below, it is more natural to formulate ${}_{ip}\dot{S}$ as a function of flux instead.

### 6.2 Formulation of internal processes

As discussed above, irreversible thermodynamics concerns the formulation of internal processes. Based on the second law of thermodynamics, all kinetics processes in a nonequilibrium system are irreversible and contribute to the total entropy change as shown by $d_{ip}S$ in Eq. 7 and Eq. 9.

For multiple independent internal processes, each can be represented by the last part of Eq. 9 and contributes to the internal energy of the system as follows for $m$ independent internal processes

$$dU = \sum_{a=1}^{n} Y^a dX^a - \sum_{b=1}^{m} D_b d\xi_b \qquad Eq.\ 47$$

$$D_b d\xi_b \geq 0 \qquad Eq.\ 48$$

where $D_b$ and $\xi_b$ are a pair of conjugate variables in analogy to $Y^a$ and $X^a$, denoted by $Y_a$ and with $X_a$ for internal variables as used in Section 3. For example, considering the diffusion of component $i$ as an internal process, i.e., $d\xi = dc_i$ with $c_i = N_i/V$ being the concentration of component $i$ in terms of moles per volume, the driving force is the decrease of the chemical potential of the component, i.e., $D = -\Delta\mu_i$. This is the same as $\mu_i$ and $dN_i$ in the combined law. With the internal energy as a function of $X^a$ and $\xi_b$, i.e., $U(X^a, \xi_b)$, the dependent variables $Y^a$ and $D_b$ are also the functions of those independent variables, i.e., $Y^a(X^a, \xi_b)$ and $D_b(X^a, \xi_b)$, respectively. When free energy functions are introduced with some molar quantities, $X^c$, replaced by their conjugate potentials, $Y^c$, the dependent potentials will follow the same as $Y^a(X^a, Y^c, \xi_b)$ and $D_b(X^a, Y^c, \xi_b)$, respectively.

When an internal process involves the change of several molar quantities *simultaneously*, such as chemical reactions involving several components, the entropy production of the internal process can be written as the sum of the entropy production of the change of each component. For a generic chemical reaction written as

$$\sum_{r} N_{A_r} A_r = \sum_{p} N_{B_p} B_p \qquad Eq.\ 49$$

with $N_r$ moles of reactant $A_r$ and $N_p$ moles of product $B_p$, its entropy production can be written as follows

$$Dd\xi = -\left(\sum_p N_{B_p}\mu_{B_p} - \sum_r N_{A_r}\mu_{A_r}\right)d\xi \qquad Eq.\ 50$$

where $d\xi$ represents the degree of the chemical reaction, i.e., $\left(\sum_r N_{A_r}\right)d\xi$ moles of reactants forming $\left(\sum_p N_p\right)d\xi$ moles of products. The driving force $D$ for the chemical reaction equals to the minus of weighted sum of products' chemical potentials subtracted by the weighted sum of the reactants' chemical potentials. It should be noted that there are internal variables that are related to microstructure such as grain size [86] and phase morphologies [87], which can be formulated into the combined law of thermodynamics with additional internal variables, but will not be discussed further in the present article because they do not directly appear in the combined law of thermodynamics.

### 6.3 Formulation of flux equations

One important type of internal processes is the transport of the molar quantity between neighboring positions in a system, i.e., its flux. Onsager [88,89] phenomenologically formulated a set of flux equations for each molar quantity by postulating its linear dependence on the gradients of all potentials and articulated that the matrix of the linear coefficients is symmetrical, commonly referred to as the Onsager reciprocal relationship. Prigogine and co-workers further developed the flux equation by including chemical reactions [90–92] and presented a unified formulation of dynamics and thermodynamics through equations of motion in terms of dynamics of correlations and instability in dissipative systems [93–95].

As discussed by the present author recently [16], there are four fundamental questions concerning Onsager's phenomenological theorem as follows

1. As a symmetric matrix can be diagonalized to obtain its eigen values (kinetic coefficients) and the eigen vector (the set of independent driving forces), what is the eigen vector?
2. When $D_b = 0$, $d\xi_b$ may not be zero because the Onsager flux equation relates $d\xi_b$ to driving forces for all internal processes, indicating that the internal processes are not truly independent.
3. The entropy production for the above scenario is zero as shown by Eq. 48, i.e., the flux of $d\xi_b$ does not produce entropy, which is in conflict with the second law of thermodynamics since any internal processes are irreversible and must result in a positive entropy production.
4. If the microscopic reversibility and Gibbs combined law hold locally, so does the Gibbs-Duhem equation as shown by Eq. 22, signifying that not all potentials or their gradients could be varied independently, and at least one of them must be a molar quantity.

Some of the above issues have been discussed in the literature as detailed by the author recently [16], particularly those by Truesdell and co-workers [96,97]. From the combined law of thermodynamics by Eq. 7, it is evident that only the product of each pair of conjugate variables enters the equation, and there are no cross-terms between non-conjugate pairs. The volumetric entropy production rate of an internal process, $d\xi_b$, can thus be written as

$$\frac{T}{\Delta V}\frac{d_{ip}S}{dt} = \frac{D_b}{A\Delta z}\frac{d\xi_b}{dt} = -\frac{\Delta Y_b}{\Delta z}\frac{\dot{X}_b}{A} = -\nabla Y_b J_{X_b} \qquad \textit{Eq. 51}$$

where $\Delta V$ is the volume of the transport with area $(A)$ and $\Delta z$ the distance between neighboring sites ($\Delta z$), $Y_b$ and $X_b$ are a pair of conjugate variables with the subscript denoting the internal variables inside the system as mentioned in Section 3, and $J_{X_b}$ is the flux of $X_b$. Consequently, in accordance with the combine law of thermodynamics, the change of a molar quantity is solely controlled by its conjugate potential gradient, and the flux equation of an internal process must be presented as follows [16]

$$J_{X_b} = -L_{X_b}\nabla Y_b \qquad \textit{Eq. 52}$$

where $L_{X_b}$ is the kinetic coefficient for the transport of $X_b$ with $-\nabla Y_b$ as its driving force. The entropy production can thus be written as

$$\sigma = {}_{\iota}\dot{S} = {}_{\iota p}\dot{S} = L_{X_b}(\nabla Y_b)^2 \geq 0 \qquad \textit{Eq. 53}$$

As discussed in details by the present author [16], Eq. 51 to Eq. 53 resolve the four questions to the Onsager flux equations presented above. For a complex internal process involving more than one $X_b$ represented in the combined law, its driving force is similarly combined by their conjugate potentials, such as the chemical reactions discussed in Section 6.2 above. In dissipative systems [93,94], the transport and chemical reactions take place simultaneously.

### 6.4 Theory of cross phenomena

The central feature that the Onsager flux equations aimed to represent is the cross phenomena [88], i.e., the experimental observations of transport of a molar quantity driven by the non-conjugate potentials such as migration of electrons by temperature gradient (thermoelectric, Seebeck coefficient) or stress gradient (electromechanical effect), or diffusion of atoms or molecules by temperature gradient (thermodiffusion, Soret coefficient). However, these observations do not provide information on the microscopic characteristics of underline physics. This is similar to the case of the Fick's first law of diffusion that correlates the atomic diffusion to its concentration gradient, but fails to represent the uphill diffusion where atoms migrate from low concentration regions to high concentration regions [16,98,99]. The diffusion of a component is driven by its chemical potential gradient which is not only a function of its own concentration gradient, but the concentration gradients or chemical potentials of all other elements in the system [16,100–103].

It is realized that an internal process can be affected by all independent variables in the system because its driving force is a function of all those independent variables as shown by the discussion in relation to Eq. 47 and Eq. 48. However, the unit process of the internal process remains the same, i.e., the migration of a molar quantity over a barrier from one state to the next state either in terms of neighboring locations such as diffusion or different structures such as chemical reactions or phase transitions. It is noted that the kinetic coefficient, i.e., $L_{X_b}$ in Eq. 52, is also a function of all those Independent variables, i.e., $Y_{X_b}(X_b, X_a, Y_c)$ and $L_{X_b}(X_b, X_a, Y_c)$ with $X_a$ and $Y_c$ denoting the other independent molar quantity and potentials. Consequently, the gradient of $Y_{X_b}$ can be written as the gradients of other independent variables as follows

$$\nabla Y_{X_b}(X_b, X_a, Y_c) = \frac{\partial Y_b}{\partial X_b}\nabla X_b + \sum\nolimits_{a\neq b}\frac{\partial Y_b}{\partial X_a}\nabla X_a + \sum\nolimits_{c\neq a\neq b}\frac{\partial Y_b}{\partial Y_c}\nabla Y_c \qquad \textit{Eq. 54}$$

Eq. 52 is then written as

$$J_{X_b} = -L_{X_b}\left(\frac{\partial Y_b}{\partial X_b}\nabla X_b + \sum\nolimits_{a\neq b}\frac{\partial Y_b}{\partial X_a}\nabla X_a + \sum\nolimits_{c\neq a\neq b}\frac{\partial Y_b}{\partial Y_c}\nabla Y_c\right) \qquad \textit{Eq. 55}$$

Eq. 55 demonstrates the cross phenomena. It overcomes the four shortcomings of the Onsager theorem discussed above and represents the fundamentals of cross phenomena. The first three shortcomings are resolved by Eq. 51 to Eq. 53, while the fourth shortcoming is addressed by Eq. 54 and Eq. 55, showing that at least one of the independent gradient is that of a molar quantity, i.e., the conjugate molar quantity of the potential. For example, in a system initially with $\nabla X_b = 0$, there is a driving force for $X_b$ to migrate due to the gradients of $\nabla X_a$ and $\nabla Y_c$ as they induce a non-zero value of $\nabla Y_b$. The flow of $X_b$ tends to transport $X_b$ in the opposite direction of $\nabla Y_b$. If the system is closed with respect to $X_b$, i.e., no exchange of $X_b$ with the surroundings, eventually the three terms on the right-hand side of Eq. 55 balance each other to result in zero driving force for $X_b$ to flow, i.e., $\nabla Y_b = 0$ and $J_{X_b} = 0$, with a nonuniform $X_b$ ($\nabla X_b \neq 0$) in the system. Consequently, the values of $X_b$ are higher than that of the original value at some regions and lower at other regions in the system, i.e., a transport phenomenon against $\nabla X_b$, commonly referred to as uphill transport.

It is evident that the significances of cross phenomena depend on the signs and magnitudes of the three sets of derivatives in Eq. 55 as discussed below.

- The first set of derivatives is between two conjugate variables and is positive for a stable system, i.e., $\frac{\partial Y_b}{\partial X_b} > 0$, shown by the diagonal quantities in Table 1 [16]. It becomes zero at the limit of stability of the system or the internal process if there are more than one internal process, and its inverse diverges positively, i.e., $\frac{\partial Y_b}{\partial X_b} = +0$ and $\frac{\partial X_b}{\partial Y_b} = +\infty$.

- The second set of derivatives is between a potential and a non-conjugate molar quantity, representing many quantities from typical experimental measurements as shown by the off-diagonal quantities in Table 1 [16]. Since they are not required to be positive, they can be sometimes negative such as the negative thermal expansion observed experimentally and predicted by the zentropy theory [104,105], which will be discussed further in Section 7. Table 1 is symmetrical due to the Maxwell relation with the ones at low-left side more easily measurable experimentally and the ones at up-right side more easily predicted theoretically [16,106]. The quantities in the last row and column of the table can be related to quantum criticality as discussed by the author [16]. At the limit of stability, they may diverge negatively, i.e., $\frac{\partial Y_b}{\partial X_a} = \pm 0$ and $\frac{\partial X_a}{\partial Y_b} = \pm\infty$. This derivative provides fundamental understanding of uphill diffusion where the diffusion of a component from a low concentration region to a high concentration region is driven by the gradient of another component, such as the carbon diffusion ($\mu_C$) driven by silicon ($c_{Si}$) as reported by Darken [16,98,99] due to the large negative value of $\frac{\partial \mu_C}{\partial c_{Si}}$.
- The third set of derivatives is between two potentials. While the second set of derivatives discussed above can be considered as cross phenomena, this third set represents the commonly referred cross phenomena in the literature where an externally controlled potential gradient, such as temperature and stress gradients and electric and magnetic fields, results in an internal $\nabla Y_c$ that drives the flows of $X_c$ and other molar quantities inside the system. These derivatives are listed in Table 2 [16,107]. Since they are not commonly seen in the literature, further discussions will be presented below.

*Table 1: Physical quantities related to the first directives of molar quantities (first column) to potentials (first row), symmetrical due to the Maxwell relations [16,106].*

*Table 2: Cross phenomenon coefficients represented by derivatives between potentials, symmetrical due to the Maxwell relations [16,107].*

As shown above, the derivatives between potentials play a central role in understanding the observation of typical cross phenomena. Their missing in the literature is probably partially because the scientific community is deeply rooted by the Onsager's phenomenological flux equation and the Onsager reciprocal relationship that stipulate the kinetic characteristics of cross phenomena and partially due to the lack of data to evaluate them. Furthermore, direct experimental measurements or kinetic simulations of cross phenomena give the products of kinetic coefficient and the derivative, i.e., $L_{X_b}\frac{\partial Y_b}{\partial Y_c}$ or $L_{X_b}\frac{\partial Y_b}{\partial X_a}$ in Eq. 55, which indeed contains the kinetic coefficient of the internal process, but not a new kinetic coefficient.

In the interest of further discussion of the kinetic coefficient matrix, let us consider the migrations of only two molar quantities using Eq. 55 as follows,

$$J_{X_b} = -L_{X_b}\left(\frac{\partial Y_b}{\partial X_b}\nabla X_b + \frac{\partial Y_b}{\partial Y_c}\nabla Y_c\right) \qquad \textit{Eq. 56}$$

$$J_{X_c} = -L_{X_c}\left(\frac{\partial Y_c}{\partial Y_b}\nabla Y_b + \frac{\partial Y_c}{\partial X_c}\nabla X_c\right) \qquad \textit{Eq. 57}$$

It is evident that the kinetic coefficient matrix is not symmetrical in general as $L_{X_b}$ and $L_{X_c}$ are independent of each other, i.e.,

$$L_{X_b}\frac{\partial Y_b}{\partial Y_c} \neq L_{X_c}\frac{\partial Y_c}{\partial Y_b} \qquad \textit{Eq. 58}$$

Alternatively, the flux equations can be written as

$$J_{X_b} = -L_{X_b}\left(\frac{\partial Y_b}{\partial X_b}\nabla X_b + \frac{\partial Y_b}{\partial X_c}\nabla X_c\right) \qquad \textit{Eq. 59}$$

$$J_{X_c} = -L_{X_c}\left(\frac{\partial Y_c}{\partial X_b}\nabla X_b + \frac{\partial Y_c}{\partial X_c}\nabla X_c\right) \qquad \textit{Eq. 60}$$

Again, one obtains the following inequality,

$$L_{X_b}\frac{\partial Y_b}{\partial X_c} \neq L_{X_c}\frac{\partial Y_c}{\partial X_b} \qquad \textit{Eq. 61}$$

This was pointed by the present author in a recently published comment [102]. It should be mentioned that the two sides in Eq. 58 or Eq. 61 could be made equal if $L_{X_b}$ and $L_{X_c}$ are not independent from each other, which means that $X_b$ and $X_c$ are not independent variables and should thus be combined to form a new independent variable. This is the case for diffusions in ionic systems where the charge neutrality constrains the equality of Eq. 58 and Eq. 61 as discussed in ref. [102] and references cited therein. This was also the case that Onsager focused on as discussed in his Nobel lecture [108] though not true in general.

To validate our above theory of cross phenomena, we calculated the electronic Helmholtz energy as a function of temperature using the Mermin formula [35] as shown in the work by Kohn and Sham [6] and Wang et al. [36] and predicted the Seebeck coefficients for a number of thermoelectric materials, showing remarkable agreement with experimental measurements [109,110]. In typical experiments, one starts with an initially homogeneous system without gradients for any molar quantities or potentials, i.e., $\nabla c_e = 0$, $\nabla \mu_e = 0$, $\nabla S = 0$, and $\nabla T = 0$, where $e$ denotes electron. When a small external temperature gradient, i.e., $\nabla Y^c = \nabla T$, is imposed to the system, it induces a heat conduction in the system and results in an internal temperature gradient in the system, i.e., $\nabla Y_c = \nabla T$. It is noted that in principles that this process takes some time to establish the internal temperature gradient, which is one of the topics that the extended irreversible thermodynamics aims to address, i.e., relaxation time [84,85]. Due to the temperature difference, the chemical potentials of electrons become inhomogeneous in the system, i.e., $\nabla \mu_e \neq 0$, inducing a driving force for electrons to migrate from high chemical potential regions to low chemical potential regions and resulting in an inhomogeneous distribution of electrons, i.e., $\nabla c_e \neq 0$. In typical experiments, the two ends of the system are not connected so electrons do not leave the system. The redistribution of electrons thus decreases $\nabla \mu_e$ until it becomes zero, i.e., $\nabla \mu_e = 0$, while the inhomogeneous electron distribution with $\nabla c_e \neq 0$ results in an internal voltage, $\nabla V_e$. The ratio of $\nabla V_e / \nabla T$ gives the experimentally measured Seebeck coefficient, noting that $\nabla V_e$ and $\nabla T$ have different signs so the Seebeck coefficient is negative for n-type thermoelectric materials. For p-type thermoelectric materials, the voltage due to the gradient of holes has the same sign as $\nabla T$, giving positive Seebeck coefficient.

In typical computer simulations to evaluate Seebeck coefficients, the heat and electron fluxes are measured by imposing either temperature gradient or electric field, and the evaluation of Seebeck coefficient thus requires the simultaneous estimation of electrical and thermal conductivity coefficients as mentioned above. Our theory of cross phenomena and computational approach avoids the evaluations of electrical and thermal conductivity coefficients in predicting the Seebeck coefficients, demonstrating better agreement with experiment measurements [109,110]. It is noted that the calculations of electrical and thermal conductivity coefficients are still needed in order to get the diagonal terms of the kinetic coefficient matrix including atomic mobilities for mass transport [109,111,112]. Furthermore, one possible next step is to combine internal processes of both transport and chemical reactions to understand and predict the properties of dissipative systems involving critical phenomena or bifurcations [95].

### 6.5 Maxwell–Stefan diffusion equation

In the Maxwell-Stefan diffusion equation [113,114], the chemical potential gradient of a component is expanded through the Maxwell–Stefan diffusion coefficients, flux, and concentration of all diffusion components as follows

$$\frac{\nabla\mu_i}{RT} = \sum_{j=1,\neq i}^{c} \frac{c_j}{c_t Đ_{ij}}\left(\frac{J_j}{c_j} - \frac{J_i}{c_i}\right) = \frac{1}{c_t}\sum_{j=1,\neq i}^{c} \frac{1}{Đ_{ij}}\left(J_j - \frac{c_j}{c_i}J_i\right) \qquad \textit{Eq. 62}$$

where $c_t$ is the total molar concentration, and $Đ_{ij}$ the Maxwell–Stefan diffusion coefficient. Complex relations between Maxwell–Stefan and commonly used diffusion coefficients have been worked out in the literature with a number of approximations for multi-component systems, relying on the Onsager reciprocal relation [113,114]. It seems that the Maxwell–Stefan approach reduces the total number of diffusion coefficients as $Đ_{ii}$ is not needed, but the Maxwell–Stefan diffusion coefficients are not based on measurable quantities and have to be estimated [114].

On the other hand, as shown by Eq. 52, one only needs one kinetic coefficient for diffusion of component $i$, i.e., $L_i$, which is in the lattice-fixed frame of reference. As presented elegantly by Andersson and Ågren [100], $L_i$ can be related to the atomic mobility ($M_i$) through diffusion mechanisms and further to the tracer diffusivity ($D_i^*$) through the Einstein relation. By changing the chemical potential gradient to concentration gradients in the lattice-fixed frame of reference, one obtains the intrinsic diffusivity (${}^iD_{ik} = L_i \frac{\partial \mu_i}{\partial c_k}$) of component $i$ with respect to the concentration gradient of component $k$, thus no longer diagomal. It is noted that the chemical potential of a component depends on the concentrations of all components where $\frac{\partial \mu_i}{\partial c_k}$ is commonly referred to as the thermodynamic factor.

On the other hand, if one switches to the volume-fixed frame of reference considering the lattice drifting due to transport of vacancy, one obtains a complex non-diagonal kinetic coefficient matrix $L'_{ij}$ so that the flux $J'_i$ in the volume-fixed frame of reference depends on the chemical potential of all components as shown by Andersson and Ågren [100]. When the chemical potential gradients are further changed into concentration gradients, one obtains the chemical diffusivity of component $i$ with respect to the concentration gradient of component $k$ ($D_{ik} = \sum_j L'_{ij} \frac{\partial \mu_j}{\partial c_k}$). The relationships among these diffusion coefficients and kinetic coefficients are shown in Figure 2 where ${}^iD_{ik}^n$ and $D_{ik}^n$ are the reduced intrinsic and chemical diffusivity of component $i$ with respect to the concentration gradient of component $k$ with component $n$ as the dependent component [16,100]. $D_{ik}^n$ can be evaluated through concentration profiles in diffusion experiments, while the evaluation of ${}^iD_{ik}^n$ needs the information on inert marker movement.

*Figure 2: Relationships among tracer diffusivity $(D_i^*)$, atomic mobility $(M_i)$, kinetic parameters $(L_i)$, and intrinsic diffusivities $({}^iD_{ik}$ and ${}^iD_{ik}^n)$ in the lattice-fixed frame of reference, and kinetic parameters $(L'_{ij})$ and chemical diffusivities $(D_{ik}$ and $D_{ik}^n)$ in the volume-fixed frame of reference* [16].

The above discussion indicates that the number of independent diffusion kinetic coefficients equals to the number of independent diffusion components. The complexity in the Maxwell-Stefan diffusion equation and its diffusion coefficients thus seems redundant as they are related to the atomic mobility and thermodynamic factors. Large scale atomic mobility databases have been developed in terms of the method shown in Figure 2 by Andersson and Ågren [100] and broadly used in diffusion simulations [115,116].

# 7 Zentropy theory for coarse graining of entropy

## 7.1 Overview of zentropy theory

It is evident from the above discussions that all theories require accurate free energy as a function of both internal and external variables. The key challenge in theoretical predictions of free energy of a phase is because only one or a few configurations are typically considered in computational approaches, whereas experimental measurements stem from sampling all possible configurations at all scales simultaneously. This challenge becomes acute for systems with phase transitions, which is often where the most fascinating transformative properties exist. This is reflected by the following gaps between statistical mechanics and quantum mechanics out of the discussions presented in Sections 4 and 5

i. Typical DFT-based calculations are focused on the ground-state configuration of a system, while experimental observations include both ground-state and non-ground-state configurations through statistical mechanics. It is thus not surprising that in general DFT-based calculations are not able to show good quantitative agreements with the experimental observations obtained at high temperature.

ii. The total energy of each configuration in statistical mechanics in Eq. 31 to Eq. 33 should be represented by Eq. 38 in quantum mechanics. However, many DFT-based calculations are performed at 0 K and thus provide only $E_k^0$ which is only part of the total energy of a configuration.

iii. The ground-state and non-ground-state configurations have non-zero entropies as shown by Eq. 39 and are thus not pure quantum states, rather coarse-grained representations of multi-body interactions of electrons and phonons for each configuration. They are thus incompatible with the existing statistical mechanics shown by Eq. 31 to Eq. 33.

The first gap has been extensively addressed in the literature through the development of an effective Hamiltonian fitted to DFT-based calculations of ground-state and some non-ground-state configurations [117–121] followed by molecular dynamics (MD) and Monte Carlo (MC) simulations. There are also approaches that directly couple DFT with MD and MC such as *ab initio* molecular dynamics (AIMD) [122–124] and quantum Monte Carlo (QMC) [125–129]. All those simulations try to sample as many configurations as possible and present the properties of a system by averaging properties of a set of well converged configurations. The selections of model, truncation, and parameter fitting in the effective Hamiltonian approach limit the quantitative predicative capabilities of MD and MC simulations in addition to the usual use of DFT data from 0 K in the fitting. Another challenge is to ensure that all important configurations are sampled in the MD and MC simulations, particularly those symmetry-breaking configurations.

The second and third gaps above are related and are not widely addressed in the literature. Ceder [130] presented a formula by replacing the total energy of each configuration in Eq. 31 by its free energy as follows

$$Z = e^{-\frac{F}{k_B T}} = \sum_{k=1}^{m} Z^k = \sum_{k=1}^{m} e^{-\frac{F^k}{k_B T}} \qquad \textit{Eq. 63}$$

with $E^k$ in Eq. 31 replaced by $F^k$. Asta et al. [131] further discussed the formula and emphasized the importance to include the entropy contribution in fitting the cluster expansion coefficients by demonstrating its effects on phase diagrams obtained from the cluster variation method. The formula was later termed as "coarse graining of the partition function" [132,133], though no actual calculations were reported in the literature using the formula by those authors.

The author's group [134,135] used the same formula of the partition function, i.e., Eq. 63, without knowing its existence at that time. We predicted the magnetic phase transition of Ce and the critical point in its temperature-pressure phase diagram, initially with two configurations, i.e., the ground-state nonmagnetic (NM) configuration and high temperature non-ground-state ferromagnetic (FM) configuration plus a mean-field term accounting for spin flipping entropy [134]. In the paper followed [135], one additional antiferromagnetic (AFM) configuration was added which removed the need of the mean-field term. The free energies of the configurations were obtained from DFT-based calculations using GGA+U.

When applying the approach to predict the negative thermal expansion in $Fe_3Pt$, the ergodicity of spin configurations in a 12 atom supercell with 9 Fe atoms was considered, resulting in $2^9 = 512$ configurations with 37 being symmetrically distinct [136]. The Helmholtz energies of all configurations were obtained by the DFT-based calculations using GGA without the need of +U, probably due to the ergodicity of configurations. Those Helmholtz energies are used to predict the critical phenomena and the negative thermal expansion along with its negative divergence at the critical point without additional models and fitting parameters. The applications to other materials were successfully performed subsequently [104,137–142] plus $YNiO_3$ with strongly correlated physics [143] and ferroelectric $PbTiO_3$ [144]. Remarkable agreement with experimental results has been observed with the first-order transitions obtained by free energy minimization and the second-order transitions defined by several criteria including the probability of the ground-state configuration decreasing to 50%, the peak of heat capacity due to statistical mixing, and more recently the inflection point in a disordering parameter [145]. It was shown that the first and last criteria give better agreement with experiments than the second one with the heat capacity.

Wentzcovitch's group [146,147] worked along the same direction with initially two spin states of Fe in $Mg_{1-x}Fe_xO$ plus a mean-field term [146], and then the order–disorder phase boundary between ice VII and VIII without the mean-field term [147]. In the latter case, the ergodicity of polar configurations was considered with 8100 configurations for 16-molecule ice VII supercell consisting of 52 symmetrically distinct configurations, noting that Ice VII is hydrogen-disordered and paraelectric, and Ice VIII is hydrogen-ordered and antiferroelectric. They defined the order-disorder transition by the peak of heat capacity and successfully applied this approach to a range of materials [148–153]. Recently the author learned that Allan and co-workers used similar approaches to study the properties of solid solutions in a number of materials at finite temperatures with each configuration having the same composition of the system [154,155]. Particularly, the Gibbs energy function, i.e., Eq. 2 in both their publications [154,155] is identical to Eq. 7 in the author's publication [156] though the entropy of mixing among configurations was explicitly presented in the latter [156] and used in the more recent formulation of the zentropy theory as discussed in detail in next section [16,105].

The success of the zentropy approach resides on the coarse graining of entropy. Bottom-up coarse graining studies the microscopic origins underlying macroscopic processes and has been broadly discussed in the scientific community [157], particularly in terms of the multiscale entropy for quantifying the complexity of physiologic time series [158,159]. It is also noted that Šafránek et al. [160] extended classical coarse-grained entropy to quantum mechanics and showed that the coarse graining using local energy measurements leads to an entropy in accord with the thermodynamic entropy. However, due to the enormous complexity of atomistic systems, statistical mechanics-driven coarse graining modeling has been very limited. The zentropy approach circumscribe this complexity by relying on the DFT-based calculations of ergodic ground- and non-ground-state configurations with the thermal electronical and vibrational entropy contributions shown by Eq. 39 and discussed in detail in the next section.

## 7.2 Fundamentals of zentropy theory: Coarse graining of entropy

The term "zentropy" was recently suggested to represent the approach by the author's group [105]. In the zentropy theory, the coarse graining of entropy is presented as follows [135]

$$S = \sum_{k=1}^{m} p^k S^k - k_B \sum_{k=1}^{m} p^k \, lnp^k \qquad \textit{Eq. 64}$$

The entropy of configuration $k$, $S^k$, can be further extended into its lower-scale configurations with the same type of formula as Eq. 64, until the electronic configurations at the DFT scale, which thus represents the bottom-up approach to compute the entropy of a configuration at the scale of the observation. While the Gibbs entropy, i.e., the second term in Eq. 64, counts the configurations from the viewpoint of the system, i.e., a top-down approach, to compute the entropy between configurations. It is thus evident that the zentropy theory integrates the bottom-up and top-down approaches as schematically shown in Figure 3 and is capable of accurately predicting the total entropy of the system, i.e., $S = \int_0^T \frac{C}{T} dT$ with $C$ being the heat capacity measured experimentally that represents a statistical sampling of ergodic configurations of the system.

*Figure 3: Schematic top-down and bottom-up integration of the zentropy theory.*

Consequently, the standard statistical mechanics in terms of Eq. 31 to Eq. 33 are modified to Eq. 63 plus Eq. 64 and the equations below

$$F = \sum_{k=1}^{m} p^k E^k - TS = \sum_{k=1}^{m} p^k F^k - k_B T \sum_{k=1}^{m} p^k lnp^k \qquad \textit{Eq. 65}$$

$$Z^k = e^{-\frac{F^k}{k_B T}} \qquad \textit{Eq. 66}$$

$$p^k = \frac{Z^k}{Z} = e^{-\frac{F^k - F}{k_B T}} \qquad \textit{Eq. 67}$$

These nested formula have several key features as follows [22]

1. Reach to the quantum regime by starting with the ground-state configuration of a system.
2. Expand to ergodic non-ground-state configurations through sampling internal degrees of freedom of the ground-state configuration.
3. Predict the entropies and Helmholtz energies of all configurations using the DFT-based calculations under the canonical ($NVT$) ensemble for each configuration.
4. Predict observable quantities with information solely from quantum mechanics through partition function of ground-state and stable non-ground-state configurations using their individual Helmholtz energies without additional models and fitting parameters.
5. Predict the free energy landscape of a system consisting of stable states, instability, critical phenomena, and free energy barriers between stable states as a function of internal and external variables,
6. Potentially extensible to many more observable quantities at various scales [143,161].

As demonstrated recently, the AFM to paramagnetic (PM) transition in $YNiO_3$ under ambient pressure was predicted to be 144 K and 81 K with or without entropy contribution for each configuration, respectively, i.e., Eq. 63 to Eq. 67 (zentropy theory) for the former and Eq. 31 to Eq. 33 (Gibbs entropy) for the latter. With the experimentally measured AFM to PM transition temperature being 145 K [143], this demonstrates the superiority of the zentropy theory. On the other hand, if the entropies of all configurations are identical, the probability of a configuration by Eq. 67 based on the zentropy theory reduces to the same formula by Eq. 33 in terms of the standard statistics due to the following relation

$$\begin{aligned} F^k - F &= E^k - TS^k - \sum_{l=1}^{m} p^l \left(E^l - TS^l + k_B T ln p^l\right) \\ &= E^k - \left(\sum_{l=1}^{m} p^l \left(E^l + k_B T ln p^l\right)\right) \end{aligned} \qquad \textit{Eq. 68}$$

However, the Helmholtz energy of the system by the zentropy theory (Eq. 65) is more negative than that by standard statistical mechanics (Eq. 32) with the difference being $-TS^k$.

There is one important point related to feature 3 and 5 above that has not been explicitly stated in the previous publications. It is easy to understand that all configurations must have the same temperature and composition, i.e., $T$ and $N$ or $N_i$, and it may be less clear that each configuration must also have the same volume because the system is in the $NVT$ ensemble. Under given $T$ and $N_i$, the Helmholtz energy of each configuration is a function of volume, and there is a volume corresponding to the lowest Helmholtz energy for each configuration. However, when individual configurations are brought together statistically, all of them must adopt the volume of the system through either compression or expansion, i.e., their Helmholtz energies need to be evaluated at the same volume as that of the system. This enables the prediction of the Helmholtz energy of transitory states between two stable states, including the inflection points that represent the limit of stability and the apex of the energy landscape. The equilibrium volume of the system is obtained through the minimization of the Helmholtz energy of the system, resulting in a single-phase state above the critical point and the mixture of two- or more-phase states below the critical point where each state has its own specific statistic mixture of all configurations, commonly referred as miscibility gap [15,105].

### 7.3 Connection between zentropy theory and entropy of black holes

The applications of zentropy theory mentioned above include magnetic and polar materials. In two overview articles [15,22], the author connected the zentropy theory to information entropy. In a recent perspective article [16], the author discussed quantum criticality, superconductivity, and the experimental observations related to quantum devices and the interpretation of second law of thermodynamics in the framework of zentropy theory, and the author's group is actively pursuing in-depth research in those directions [161]. Furthermore, the potential applicability of zentropy theory to the entropy of black hole [162–165] was mentioned [16,22], and some preliminary thoughts are elaborated in more details in the present section.

The entropy of black hole was formally formulated by Bekenstein [162] and reviewed by Hawking [163] and commonly referred as Bekenstein-Hawking entropy in the literature. Bekenstein [162] presented the law of black holes in terms of the changes of internal energy, $d(Mc^2)$, area, $dA$, angular momentum, $dJ$, and charge, $dQ$, as follows

$$d(Mc^2) = \frac{\kappa c^2}{8\pi G} dA + \Omega dJ + \Phi dQ \qquad \textit{Eq. 69}$$

where $M$, $\kappa$ and $\Omega$ are the mass, surface gravity and the angular frequency of rotation of the black hole, $c$ is the speed of light, $G$ is the gravitational constant, and $\Phi$ is the potential of the event horizon, i.e., the boundary between the black hole and outside world. Bekenstein compared Eq. 69 with Eq. 13, i.e., the Gibbs combined law of thermodynamics and identified the last two terms in Eq. 69 as the work done on the black hole by an external agent who increases the black hole's angular momentum and charge by $dJ$ and $dQ$, respectively, in analog of $-PdV$ in Eq. 13 or more general $dW$ in Eq. 7.

A critical step next was to correlate the area and entropy between Eq. 69 and Eq. 13. Bekenstein [162] started from the information entropy as shown by Eq. 30, forgo the internal configurations of a black hole due to their inaccessibility, and derived the entropy of a black hole as a linear function of area as follows

$$\frac{S_{bh}}{k_B} = \gamma \frac{c^3}{\hbar G} A \qquad \textit{Eq. 70}$$

where $\hbar$ is the reduced Planck constant, and $\gamma = \frac{ln2}{8\pi}$ was obtained by Bekenstein [162] and $\gamma = \frac{1}{4}$ by Hawking [163], respectively. It is noted that Bekenstein [162] emphasized that "the concept of black-hole entropy as the measure of the inaccessibility of information (to an exterior observer) as to which particular internal configuration of the black hole is actually realized in a given case", referring to the "equivalence class of all black holes which have the same mass, charge, and angular momentum, rather than the thermal entropy inside the black hole".

In a following paper, Bekenstein [166] presented a detailed statistical analysis in terms of the statistical sampling of a black hole. Hawking [163] discussed the quantum fluctuation of internal configurations or quantum states of a black hole and concluded that a black hole radiation can take place by statistical fluctuations in black-body radiation and can then decay quantum mechanically with the reemission of radiation. Over the years, many microscopic explanations of black hole entropy have been developed through statistical mechanics as reviewed by Carlip [167,168], including entanglement entropy, string theory, loop quantum gravity, induced gravity, and logarithmic corrections. The central challenge is to define and count the internal configurations of black holes.

There have been some revisions of the original derivation by Bekenstein [162] with one being the addition of a pressure term to the law of black holes [169–174] by treating the cosmological constant as a thermodynamic pressure and the other being the logarithmic corrections of entropy formula by considering the effect of thermal fluctuations [175–183]. It should be noted that the Gibbs combined law of thermodynamics was stated as the first law of thermodynamics by Hawking and many others in the literature, which should be the combined law as shown in Sections 2.3 and 3.1. This may be partially due to the fact that the second law of thermodynamics was removed from Gibbs combined law of thermodynamics as discussed in Sections 2.3 and 3.1. Furthermore, the Gibbs combined law of thermodynamics used by Bekenstein [162] and Hawking [163] is for closed systems, while a typical *gedanken* experiment involves the falling of a box, Wheeler's cup of tea [167], or other items into a black hole. Consequently, it is necessary to use the combined law including the internal processes and all exchanges between a black hole and its surroundings, i.e., Eq. 7 for the combined law of thermodynamics and Eq. 9 for the entropy production due to internal processes.

Combining Eq. 69 and Eq. 7 with the two revisions of pressure and mass exchange mentioned above results in the following equation similar to those in the literature [169–174]

$$dU = TdS + \Omega dJ + \Phi dQ - PdV + \sum_{i=1}^{c} \mu_i dN_i - Td_{ip}S \qquad \textit{Eq. 71}$$

where $dN_i$ denotes the moles of component $i$ added to the black hole as no mass escapes the black hole, and $d_{ip}S$ the entropy production due to internal processes inside the black hole. It should be emphasized that $dS$ is for the entropy change of the black hole and can be either positive or negative as shown by Eq. 5, which is not directly related to the second law of thermodynamics. As discussed in Section 2.2, the second law of thermodynamics concerns only the last term of Eq. 71 as shown by Eq. 4. Therefore, there is no need to introduce the generalized second law of thermodynamics [162,163]. In Eq. 71, $-PdV$ is used instead of $VdP$ commonly presented in the literature [169–174] in order to avoid the use of enthalpy on the left side of Eq. 71 though they are equivalent as both are fundamental characteristic functions [20].

Let us first discuss the logarithmic correction to the Bekenstein-Hawking entropy (Eq. 70) that takes a typical form as follows [175–183]

$$S'_{bh} = S_{bh} + \eta ln(S_{bh}) + additional\ terms \qquad Eq.\ 72$$

with $\eta$ being a constant. On the other hand, Bekenstein [166] discussed a different statistical interpretation of the concept of black-hole entropy as the natural logarithm of the number of possible states of a black hole that are compatible with the given spinning black-hole state. Since only the spinning state of a black hole is observable, Bekenstein [166] articulated that a black hole cannot be regarded as having definite $M$, $L$, and $Q$, rather in a number of different spinning black hole solution states of definite $M$, $L$, and $Q$, each one occurring with some probability $p_{MLQ}$, identified as the sum of probabilities for all radiation states that can coexist with the given spinning solution state. The following formula of entropy of a black hole was then obtained by Bekenstein [166]

$$S'_{bh} = \sum_{MLQ} p_{MLQ} S_{bh}(M, L, Q) - k_B \sum_{MLQ} p_{p_{MLQ}} ln p_{p_{MLQ}} \qquad Eq.\ 73$$

where $S_{bh}(M, L, Q)$ is the entropy of a spinning black-hole state with the parameters $M$, $L$, and $Q$. This approach from the higher observable scale to the lower unobservable scale may be compared with the top-down approach in our multiscale entropy approach [22]. It is evident that in this approach, each $MLQ$ observation represents a specific statistical distribution of ergodic individual configurations in the black hole though they are not accessible from outside and is similar to MC and MD sampling.

Eq. 73 is strikingly similar to that of our zentropy theory shown by Eq. 64, which is also similar to that derived for entanglement entropy in relation to quantum criticality [184–190] where the area law dominates when the system is far away from its quantum critical point, and the logarithmic law dominates when the system is near its quantum critical point. Our zentropy theory was able to predict the singularity at critical points in Ce [134,135] and $Fe_3Pt$ [136,145] due to the magnetic spin dynamics in the temperature-pressure two-dimensional space including the positive divergency of thermal expansion in Ce and the negative divergency of thermal expansion in $Fe_3Pt$ [15,16,156].

It is noted that Eq. 64 and Eq. 73 are capable of interpreting the transition from the area law to the logarithmic law when a system is approaching its critical point either from its ground state or a non-ground state far away passing the critical point. At the ground state, there is only one configuration in the system, i.e., the ground-state configuration. The entropy of the system equals to the entropy of the ground state, which is a function of the size of the system either in terms of volume or surface, thus the area law. At the non-ground state far away from the critical point, the probability of each configuration in the system is same, and Eq. 64 becomes

$$S = \frac{1}{m}\sum_{k=1}^{m} S_k + k_B ln(m) \qquad \textit{Eq. 74}$$

The entropy of the system is dominated by the first term and thus the area law. While near the critical point, the ground-state configuration loses its dominance, and the system fluctuates spontaneously between the ground-state configuration and non-ground-state configurations, resulting in an inflection point on the degree of disorder derived from the entropy change as a function of temperature, and thus the logarithmic law. The system diverges at the critical point, and the fluctuation wavelength becomes infinite, even only with quantum information from supercells as small as 12 atoms for $Fe_3Pt$ [136,156]. The divergence of effective mass of electrons at the quantum critical point was discussed similarly [16]. It thus seems plausible that the zentropy theory has the potential to be applied to predict the properties of black holes, which will be explored further in our future research activities.

Concerning the addition of $-PdV$ to Eq. 71, there seem some inconsistences. As a volume change usually results in an area change, Eq. 71 is thus inconsistent with Eq. 69 since both entropy and volume are independent variables of the internal energy of a black hole, i.e., the entropy cannot be correlated with area or volume directly as Bekenstein [162] did. However, this issue has not been addressed in the literature when the pressure or volume was introduced [169–174]. On the other hand, the discussion in the above paragraph demonstrates that the area law can be rationalized through the volume or area dependence of the entropy of the ground-state configuration, or the statistical mixture of all configurations as shown by Eq. 74 without the intuitive suggestion between entropy and black-hole area made by Bekenstein [162].

Next let us correlate the Hawking radiation in the framework discussed in the present work. The Hawking radiation reduces the mass and rotational energy of a black hole through the energy radiated from the black hole to its surroundings. The entropy change of the black hole may thus be written as follows from Eq. 5,

$$dS = \frac{dQ_H}{T} + d_{ip}S \qquad \textit{Eq. 75}$$

where $dQ_H$ is the heat loss of the black hole due to the Hawking radiation (thus negative), and $d_{ip}S$ is the entropy production due to the internal process and can be written as follows from Eq. 9

$$d_{ip}S = \frac{d_{ip}Q}{T} - \sum_j S_j dN_{r,j} + d_{ip}S^{config} \qquad \textit{Eq. 76}$$

where $dN_{r,j}$ is the moles of component $j$ converted into energy represented by the heat generation $d_{ip}Q$ inside the black hole which may be approximated as $d_{ip}Q = c^2 \sum_j dN_{r,j} = -c^2 dM$ with $dM$ being the mass reduction of the black hole (thus negative), and $d_{ip}S^{config}$ is the change of the configurations inside the black hole. Considering the black hole in a relatively steady state with $d_{ip}S^{config} \approx 0$ and $c^2 > TS_j$, one has $Td_{ip}S \approx -c^2 dM > 0$, in accordance with the second law of thermodynamics. The entropy change of the black hole with Hawking radiation and the reduction of black hole mass can thus be approximated as

$$TdS \approx dQ_H - c^2 dM \qquad \textit{Eq. 77}$$

If the entropy of a black hole remains approximately constant when the heat released by the Hawking radiation is balanced by the mass to energy conversion inside the black hole, i.e.,

$$dQ_H = c^2 dM \qquad \textit{Eq. 78}$$

The mass of the black hole thus continuously decreases due to the Hawking radiation, i.e., $dQ_H < 0$ and $dM < 0$.

Concerning a *gedanken* experiment on a box, Wheeler's cup of tea [167], or other items falling into a black hole, it may be easier to consider the box and the black hole as one system, and the falling process is thus an internal process of the system and can be treated as an internal flux as discussed in Section 6.3 with the gravitational force being the driving force. The reaction after the item falls into the black hole can be either combined with the falling process as one internal process or considered as another internal process as they are independent of each other. At some stage inside the black hole, the mass of the box will convert into energy as discussed above in relation to the Hawking radiation.

### 7.4 On deterministic vs probabilistic models

Deterministic and probabilistic (or stochastic) models are usually considered as opposite to each other such as quantum mechanics vs Newtonian mechanics. In statistical mechanics represented by either Eq. 30 to Eq. 33 for pure quantum configurations or Eq. 64 to Eq. 67 for non-pure quantum configurations, the appearance of each individual configuration in the system is only a probability, indicating that the measurements would be very sporadic if conducted at the temporal and spatial resolutions that the system experiences various configurations which is dictated by the barriers between the configurations such as those between the ferroelectric configurations of $PbTiO_3$ without a domain wall or with 90 or 180 degree domain walls, respectively [144]. When the experimental resolutions are finer than those of system switching, one observes ferroelectric behaviors with a tetragonal structure and macroscopic polarization. When it is the opposite, one observes paraelectric behaviors with a cubic structure and without macroscopic polarization. This is similar to the bird cage thaumatrope where one observes bird and cage separately with a slow spin speed and bird-in-cage for a high spin speed [191]. It is thus self-evident that the behaviors of individual configurations at its own scale is stochastic, while their collective behaviors at larger scale are deterministic, which is captured by statistical mechanics discussed in the present paper, particularly the zentropy theory that nests all the way down to the scale of electrons in quantum mechanics with the Heisenberg's uncertainty principle.

In discussing the passivity of metals as the key to our metals-based civilization, Macdonald [192] articulated that the formation of a thin reaction product film on the metal surface is deterministic, and there are physical models can account for most, if not all, experimental observations and provide a robust basis for predicting the occurrence of passivity breakdown and the evolution of localized corrosion damage in a wide range of systems. It was further pointed out that the transition of a specific metastable event to a stable event, and hence the nucleation of a stable pit, is a rare probabilistic event, thus stochastic, determined by the kinetics of repassivation that is dependent on the chemical composition of the environment, the nature of the breakdown sites, and on the electrochemical properties of the system. It was also discussed that a stochastic process incorporating short-term memory effects does indeed yield the experimentally observed near-normal distributions in a critical breakdown voltage in agreement with those derived deterministically.

In discussing the computable universes, Schmidhuber [193–195] mentioned that although macroscopic properties of a system can often be predicted by physical laws, microscopic properties are subject to fluctuations, representing additional information absent in the macroscopic physical laws. It was further pointed out that true randomness in quantum mechanics means that there is no existing short algorithm that can compute the precise collapse of the wave function which could be due to the true randomness or our fundamental limited understanding of the universe. A profound question is whether there exists a very short program that can calculate the entire history and future of any systems and yield not only the known physical laws but also every single seemingly random elementary event in the systems. For efficiency and practicality, truncating the gradient through the long short-term memory (LSTM) approach may be the way to go by enforcing constant error flow through constant error carousels within special units [196].

The present paper shows the statistical nature of entropy from the quantum scale in terms of electrons and phonons to the scale of black holes. Yet the second law of thermodynamics is deterministic in dictating the positive entropy production of any internal processes, while the entropy of a system can either increase or decrease depending on the interaction between the system and its surroundings. For example, a system under internal statistical equilibrium has all its macroscopic properties well-defined, while the properties of individual atoms inside the system are truly statistical due to Brownian motion which is also governed by the second law of thermodynamics. Nevertheless, one may ask what if the system includes the whole universe or all universes [193–195]. Unfortunately, this question could not be answered because there would be no surroundings of the system, thus no observers.

This probably reflects the foundational value of the nested formula of the zentropy theory, i.e., coarse graining at the scales below observation and truncating in terms of the current limits of our knowledge of physics and the infinite number of possible configurations. Our current knowledge of quantum systems is limited by quantum mechanics which is statistical in nature and can only predict the probabilities of various possible electron distributions. However, by coarse graining of all electrons, DFT prescribes that there is one unique electron density distribution for the ground state of a given system at zero K, which further determines all observables of the system at scales higher than electrons. This coarse graining process results in a deterministic outcome from probabilistic lower-level information. By varying temperature and pressure, probabilities of metastable non-ground-state configurations become non-zero, and their statistical mixture with the ground-state configuration results in measurable deterministic outcomes though with certain uncertainty due to fluctuations of individual configurations, which is Eq. 73 proposed by Bekenstein [166]. In some systems, it can produce one or more critical points with singularity, and when such a critical point is close to zero K, one has a quantum critical point. In addition to pressure, singularity can be induced by any external variables shown in the combined law of thermodynamics (Eq. 7). Considering the similarity of Eq. 64 and Eq. 73, it seems plausible that the singularity of black holes may also be predictable by the nested formula of the zentropy theory through the deterministic and probabilistic integration.

## 8 Perspectives on future of thermodynamic modeling

In the CALPHAD method, the Gibbs energies of individual phases are modeled as a function of temperature, pressure, and composition which are controlled from the surroundings and dictate the ground-state configuration of the system, and additional internal variables that represent the non-ground-state configurations. The Gibbs energy builds from pure element to binary and ternary systems and extrapolates to multicomponent systems. In this special issue, Spencer presented overviews of its development [10], and Olson and Liu [197] discussed the computational materials design built on the CALPHAD method. In addition to the monograph by Kaufman and Bernstein [8], there are two books [198,199] dedicated to and other books [20,21,200] discussed the CALPHAD method. The applications of CALPHAD databases have been addressed in many other publications [10,201] including those by the present author [15,197].

The author's perspectives on the future of thermodynamic modeling are discussed in this section in three areas: the foundational lattice stability in the CALPHAD modeling, the highly accurate free energy data of individual phases and their efficient generation to circumvent the lack of phase equilibrium data for new materials, and the development of new tools for more automated procedures and new models to include contributions from external electric and magnetic fields and internal variables such as defects.

### 8.1 Lattice stability

As mentioned in the introduction, the digitization of thermodynamics of multicomponent multiphase materials has been accomplished by the CALPHAD method which models the free energy of each individual phase as a function of external and internal variables [8,15,197–199]. As a phase can be stable, metastable, or unstable under given external conditions, the CALPHAD method effectively treats the phase fraction, i.e., 100% for an individual phase, as an internal variable in addition to other internal variables in non-ground-state configurations such as short- and long-range ordering of atomic species and spontaneous magnetic and electrical polarizations. One key issue identified by Kaufman was the modeling of a solution phase of two elements with different stable structures under ambient conditions, such as the bcc solid solution phase of Fe-Ni where the free energy of bcc-Ni must be defined [7,202]. The free energy difference between the stable and nonstable crystal structures of a pure element was subsequently termed as the "lattice stability" [203–205] and enabled the free energy modeling of individual phases across the full composition space of multicomponent materials [8].

The pivotal role of the lattice stability is its inter-dependence with the interaction parameters in a solution phase. Consequently, the value of a lattice stability must be the same for all binary systems using this lattice stability, and a change of its value results in the need to revise all those binary systems and ternary and higher-order systems built on those binary systems. The currently used lattice stability of pure elements, commonly referred as SGTE91 [206], was compiled more than 30 years ago and has enabled the development of many commercial databases for multicomponent and multiphase materials [115,116,207] and their successful applications to computational materials design [10,208,209], which along with the success of the Human Genome Project [210] prompted the author to coin the term "materials genome"® denoting the individual phases as the building blocks of materials [211].

It was inevitable that the definition of lattice stability and its evaluation were challenging and heavily debated from the beginning [44,212] because of their conceptual importance in paradigm change for thermodynamic modeling and at the same time many of those nonstable structures being unstable. The limit of stability was discussed in Section 6.4 above. The entropy and thus free energies of unstable states could not be directly predicted by DFT-based calculations through Eq. 37 and Eq. 39 due to imaginary vibrational frequencies. While various theoretical approaches have been discussed in the literature [45–52], they all have individual strengths and weaknesses through various degrees of compromise between practical usefulness and physical soundness. Common to most approaches is on the extrapolation from stable regions to unstable regions in terms of independent variables such as composition, pressure, temperature, or lattice distortion.

van de Walle [49,50] showed that the energies at the limit of stability of several pure elements from DFT-based calculations agree with those in SGTE91. As noted by Grimvall [213], a binary solution becomes unstable before reaching the unstable pure element. It is intuitive to think that the extrapolation used in developing the lattice stability in SGTE91 was able to sense the limit of stability, thus the agreement between the results by van de Walle [49,50] and SGTE91. However, it is not fully satisfactory because if the pure element or the solution is quenched from a stable state to this unstable state, one would like to have the free energy of the true unstable state at various compositions in order to understand or simulate the transition from the unstable state to a stable state, including the spinodal decomposition in many binary and multicomponent systems. It is noted that Kadkhodaei, Hong, and van de Walle [214] considered the occurrence of hopping between local low-symmetry distortions for a high-symmetry time-averaged structure at high temperature. To compute the free energy in such phases, they explored the system's potential energy surface by discrete sampling of local minima through a lattice gas Monte Carlo approach and by a continuous sampling by means of a lattice dynamics approach in the vicinity of each local minimum. The bcc Ti was used as an example to illustrate the approach, which was further extended to study diffusion properties in bcc Ti and other elements [215,216]. This approach is similar to the zentropy theory in considering a phase as a mixture of local minima or configurations, but requires the information on the existence of such a phase at high temperature. At 0 K, local minima are the ground-state or non-ground-state configurations, while at high temperatures each local minimum has contributions from all configurations which reaches the extreme at the critical point between stable and unstable states of the high-symmetry time-averaged macroscopically homogeneous structure as shown in Ce and $Fe_3Pt$ [105].

On the other hand, Yang et al. [51] discussed the Cr lattice stability derived by the CALPHAD and *ab initio* approaches and concluded that the *ab initio* lattice stability of fcc-Cr at zero K can be a viable approach as demonstrated through the modeling of the Fe-Cr and Ni-Cr binary systems, though the free energy of fcc-Cr at finite temperature was not discussed. Since the zentropy theory can predict the free energies of unstable states as a function of internal variables based on the statistical competition among stable ground-state and metastable non-ground-state configurations, it is reasonable to expect that the zentropy theory may have the potential to address this challenge. Consequently, those configurations may be considered as the building blocks instead of individual phases [211]. A recent attempt was made to predict the free energies of fcc, bcc, and hcp Fe including magnetic properties [217]. The subsequent challenge is how to obtain all the configurations and potentially the astronomic number of configurations as discussed in the next section.

## 8.2 Input data for thermodynamic modeling from zentropy and machine learning models

In principles, thermodynamic modeling could be performed with only thermochemical data as they are the derivatives of free energy. However, most of thermochemical data are derived from measurements of heat with large uncertainty, and Gibbs energies of individual phases thus evaluated cannot give accurate transition conditions between phases. Consequently, the Gibbs energy model parameters of all phases need to be refined simultaneously using experimentally measured phase transition data. This refinement step not only requires additional experimental input, but also make the model parameters of all phases dependent on each other though they may be related in principle to some degree as implicated by the zentropy theory discussed above.

DFT-based calculations have provided useful input data for CALPAHD modeling [12]. However, as discussed in the present paper, each calculation is for one given configuration and does not represent the properties of the phase. While the zentropy theory has demonstrated its capability to accurately predict magnetic and ferroelectric transitions for stoichiometric phases, its applicability to solution phases remains to be tested in terms of both efficiency and accuracy. The author's group is actively developing tools and data infrastructure for the zentropy approach and its application to solution phases.

Since DFT-based calculations are both computing resource intensive and complex for free energies, the author's group has developed the open source DFT Tool Kits (DFTTK) [38,39] that streamlines the calculations of entropy and post processes to obtain free energy. In last several years, machine learning (ML) models based on deep neural networks (DNN) have been vastly implemented into the materials science community. The author's group has developed such a DNN ML model, named SIPFENN (Structure-Informed Prediction of Formation Energy), for predicting formation energy at zero K [218,219], which can be installed through PyPI by pip install pysipfenn. The author's group is actively developing DNN ML models for free energy of given configurations, providing data for the zentropy approach. Other ML models are being developed by the community such as deep-learning models using the atomistic line-graph neural network (ALIGNN) [220,221]. It is anticipated that more and more ML models will be developed in the community to predict free energy as a function of all possible internal and external variables.

Another highly desired set of data for CALPHAD modeling is the properties of the liquid phase. In principle, the properties of the liquid phase can be predicted by AIMD [122–124] including heat capacity [222], diffusivity [223], and enthalpy of mixing [224]. More recently, Hong and van de Walle [225] presented an open-source code to predict the melting temperature and enthalpy of fusion employing Born–Oppenheimer molecular dynamics techniques under the isobaric–isothermal (NPT) ensemble and their small-size liquid-solid coexistence method. This has further enabled them to develop ML models for efficient prediction of properties of liquid of a broad range of materials [226,227]. These approaches have the potential to greatly enhance the quality of multicomponent CALPHAD databases, particularly for discovery and design of new materials for high temperature applications.

One key factor in the zentropy theory is the number of configurations in materials systems. The Human Genome Project reports approximately 22,300 protein-coding genes with about 3.1 billion base pairs [210]. Their ratio is in the order of $10^5$, which is approximately equivalent to the number of configurations of a supercell with 11 lattice sites in a ternary phase, i.e., $3^{11} \approx 1.8 \cdot 10^5$, which could be lowered as some of configuration are equivalent due to the symmetry that can be checked, for example using the ATAT code [228]. These configurations cover three pure elements, 30 compositions in three binary systems, and 45 ternary compositions, total 78 compositions. In addition to DFT-based prediction for the free energy of all those configuration, the cluster expansion implemented in the ATAT code [228] poses as an efficient approach to predict the formation energy at zero K of various atomic configurations for a given lattice structure and can be extended to the prediction of their free energies [133]. Considering the 78 elements in the SGTE pure element database [206], the number of ternary systems is $C(78,3) = 76{,}076$. Assuming one third of the ternary systems being practically useful and 10 phases per ternary system, the total number of configurations is roughly estimated to be in the order of tens billions. The author hopes that the prediction of their free energy as a function of temperature, pressure/stress or strain, and electric/magnetic fields may be available in not distant future with new physics-informed NN (PINN) ML models [229]. It is further noted that recent approaches such as supercell random approximates (SCRAPs) [230] and small set of ordered structures (SSOS) [231] may help to reduce the supercell sizes and the number of supercells needed.

An potential future application of the zentropy theory and the theory of cross phenomena [16] is to predict kinetic coefficients for transition between two states. As shown for Ce [135] and $Fe_3Pt$ [136] with 9 and 512 configurations, respectively, a critical point in the temperature/pressure and temperature/volume phase diagrams was predicted by the zentropy theory. Above the critical point, there is only one minimum on the Helmholtz energy as a function of volume, and the transition is 2nd order. While below the critical point, there are two minima on the Helmholtz energy as a function of volume, and the transition is 1st order between the two states of different volumes with different mixtures of configurations. The common tangent of the two minima gives the pressure of the two-state equilibrium, and their respective volumes represent the local equilibrium condition at their interface during transition with the maximum on the Helmholtz energy denoting the kinetic barrier for the transition through either spinodal transformation for an initially unstable state between the two inflection points on the Helmholtz energy curve or nucleation and growth for an initially metastable state outside the two inflection points.

For diffusion processes, the author's team used the DFT-based method to predict the tracer diffusivity by manually moving the atom along the diffusion pathway [111]. However, when the atom crosses the inflection point on the free energy curve, and the system becomes unstable, the entropy could not be calculated due to imaginary frequencies of phonon vibration, and the transition state theory had to be used by imposing a constraint preventing from oscillating along the direction (in momentum space) parallel to the diffusion direction to effectively remove unstable vibrational modes. It would be an interesting research topic to see how the zentropy theory could be applied to this case with the position of the diffusion atom along the diffusion direction as an internal variable.

### 8.3 Models and tools for thermodynamic modeling

There are well developed commercial tools for CALPHAD modeling and a wide range of large CALPHAD databases for education, research, and industrial applications [115,116,207]. For the continuous progress of the field, there is a need for new tools so new thermodynamic and other property models can be developed, tested, and compared. Furthermore, the interdependence of model parameters in different phases mentioned above makes the improvement of modeling difficulty because change of one parameter necessitates the change of many other parameters [13,14].

The author's group started to develop an automated CALPHAD modeling tool, Extensible Self-optimizing Phase Equilibria Infrastructure (ESPEI), a while ago with limited success due to the lack of flexible computational engine [232]. Recently, the group developed an open-source software package, PyCalphad, for thermodynamic calculations [233,234] and used it to develop a complete new ESPEI code [235,236]. Furthermore, the modified quasichemical model in quadruplet approximation (MQMQA) has been implemented in the software packages [237] with a number of other models being programed including the universal quasichemical (UNIQUC) [238] model and its improved variants [239,240] and Peng-Robinson equation [241] widely used in the oil and gas community to describe critical phenomena between gas and liquid [242–244]. PyCalphad and ESPEI are open-source as part of an open source software ecosystem [245], available for scientists to implement and test their own models, and capable of uncertainty quantification [246] and sensitivity analysis [247].

## 9 Summary

Thermodynamics is at the core of science and nature, and thermodynamic modeling based on the CALPHAD method has enabled the community to quantitatively go beyond the equilibrium applications of thermodynamics, understand and improve existing materials, and design new materials. The present overview paper further discussed the zentropy theory and the theory of cross phenomena for better prediction of data for CALPHAD modeling. Based on the integration of quantum mechanics and statistical mechanics, the zentropy theory provides new capabilities to accurately predict free energy of individual phases from the DFT-based calculations starting from the ground-state and non-ground-state configurations. In keeping the entropy production due to internal processes in the combined law of thermodynamics, the theory of cross phenomena provides fundamental understanding of interactions of multi-variables and mathematical approaches to predict the cross phenomena coefficients. The author's perspectives on the potential applications of the zentropy theory and future directions of CALPHAD modeling are also presented. While the future is deterministic, the pathways can be very stochastic and uncertain and can involve singularities.

**Acknowledgements.** The present review article covers research outcomes supported by multiple funding agencies over multiple years with the most recent ones including the Endowed Dorothy Pate Enright Professorship at the Pennsylvania State University, U.S. Department of Energy (DOE) Grant No. DE-SC0023185, DE-NE0008945, and DE-NE0009288, and U.S. National Science Foundation (NSF) Grant No. NSF-2229690. The author would like to thank J.P. Perdew for explaining their recent works on symmetry breaking due to fluctuations of various wavevectors.

## 10 References

1. Sinnott, S. B. & Liu, Z. K. Predicted Advances in the Design of New Materials. *The Bridge* **50S,** 147–149 (2020).
2. Liu, Z. K. Materials 4.0 and the Materials Genome Initiative. *Adv. Mater. Process.* **178(2),** 50 (2020).
3. Schrödinger, E. Quantisierung als Eigenwertproblem. *Ann. Phys.* **384,** 361–376 (1926).
4. Schrödinger, E. An Undulatory Theory of the Mechanics of Atoms and Molecules. *Phys. Rev.* **28,** 1049–1070 (1926).
5. Hohenberg, P. & Kohn, W. Inhomogeneous electron gas. *Phys. Rev. B* **136,** B864–B871 (1964).
6. Kohn, W. & Sham, L. J. Self-Consistent Equations Including Exchange and Correlation Effects. *Phys. Rev.* **140,** A1133–A1138 (1965).
7. Kaufman, L. & Cohen, M. The Martensitic Transformation in the Iron-Nickel System. *JOM* **8,** 1393–1401 (1956).
8. Kaufman, L. & Bernstein, H. *Computer Calculation of Phase Diagrams*. (Academic Press Inc., 1970).
9. Spencer, P. J. A brief history of CALPHAD. *CALPHAD* **32,** 1–8 (2008).
10. Spencer, P. J. The origins, growth and current industrial impact of Calphad. *CALPHAD* **79,** 102489 (2022).
11. Gubernatis, J. E. & Lookman, T. Machine learning in materials design and discovery: Examples from the present and suggestions for the future. *Phys. Rev. Mater.* **2,** 120301 (2018).
12. Liu, Z. K. First-Principles calculations and CALPHAD modeling of thermodynamics. *J.*

*Phase Equilibria Diffus.* **30,** 517–534 (2009).

13. Campbell, C. E., Kattner, U. R. & Liu, Z. K. File and data repositories for Next Generation CALPHAD. *Scr. Mater.* **70,** 7–11 (2014).

14. Campbell, C. E., Kattner, U. R. & Liu, Z. K. The development of phase-based property data using the CALPHAD method and infrastructure needs. *Integr. Mater. Manuf. Innov.* **3,** 158–180 (2014).

15. Liu, Z. K. Computational thermodynamics and its applications. *Acta Mater.* **200,** 745–792 (2020).

16. Liu, Z. K. Theory of cross phenomena and their coefficients beyond Onsager theorem. *Mater. Res. Lett.* **10,** 393–439 (2022).

17. Gibbs, J. W. Graphical methods in the thermodynamics of fluids. *Trans. Connect. Acad. II* **April-May,** 309–342 (1873).

18. Gibbs, J. W. On the equilibrium of heterogeneous substances. *Am. J. Sci.* **s3-16,** 441–458 (1878).

19. Gibbs, J. W. *The collected works of J. Willard Gibbs: Vol. I Thermodynamics*. (Yale University Press, Vol. 1, 1948).

20. Hillert, M. *Phase Equilibria, Phase Diagrams and Phase Transformations*. (Cambridge University Press, 2007). doi:10.1017/CBO9780511812781

21. Liu, Z.-K. & Wang, Y. *Computational Thermodynamics of Materials*. (Cambridge University Press, 2016). doi:10.1017/CBO9781139018265

22. Liu, Z. K., Li, B. & Lin, H. Multiscale Entropy and Its Implications to Critical Phenomena, Emergent Behaviors, and Information. *J. Phase Equilibria Diffus.* **40,** 508–521 (2019).

23. In *The IUPAC Compendium of Chemical Terminology* (eds. McNaught, A. D. & Wilkinson, A.) (International Union of Pure and Applied Chemistry (IUPAC), 2008). doi:10.1351/goldbook.G02629

24. In *The IUPAC Compendium of Chemical Terminology* (eds. McNaught, A. D. & Wilkinson, A.) (International Union of Pure and Applied Chemistry (IUPAC), 2014). doi:10.1351/goldbook.H02772

25. Gibbs, J. W. *The collected works of J. Willard Gibbs: Vol. II Statistical Mechanics*. (Yale University Press, Vol. II, 1948).

26. Landau, L. D. & Lifshitz, E. M. *Statistical Physics*. (Pergamon Press Ltd., 1970).

27. Born, M. & Oppenheimer, R. Quantum theory of molecules. *Ann. Phys.* **84,** 457–484 (1927).

28. Ceperley, D. M. & Alder, B. J. Ground state of the electron gas by a stochastic method. *Phys. Rev. Lett.* **45,** 566–569 (1980).

29. Perdew, J. P. & Zunger, A. Self-Interaction Correction to Density-Functional Approximations for Many-Electron Systems. *Phys. Rev. B* **23,** 5048–5079 (1981).

30. Perdew, J. P. & Wang, Y. Accurate and simple analytic representation of the electron-gas correlation energy. *Phys. Rev. B* **45,** 13244 (1992).

31. Perdew, J. P., Burke, K. & Ernzerhof, M. Generalized Gradient Approximation Made Simple. *Phys. Rev. Lett.* **77,** 3865–3868 (1996).

32. Perdew, J. P., Ruzsinszky, A., Csonka, G. I., Vydrov, O. A., Scuseria, G. E., Constantin, L. A., Zhou, X. & Burke, K. Restoring the Density-Gradient Expansion for Exchange in Solids and Surfaces. *Phys. Rev. Lett.* **100,** 136406 (2008).

33. Kresse, G. & Joubert, D. From ultrasoft pseudopotentials to the projector augmented-wave

method. *Phys. Rev. B* **59,** 1758–1775 (1999).

34. Shang, S. L., Saengdeejing, a., Mei, Z. G., Kim, D. E., Zhang, H., Ganeshan, S., Wang, Y. & Liu, Z. K. First-principles calculations of pure elements: Equations of state and elastic stiffness constants. *Comput. Mater. Sci.* **48,** 813–826 (2010).

35. Mermin, N. D. Thermal Properties of the Inhomogeneous Electron Gas. *Phys. Rev.* **137,** A1441–A1443 (1965).

36. Wang, Y., Liu, Z.-K. & Chen, L.-Q. Thermodynamic properties of Al, Ni, NiAl, and Ni3Al from first-principles calculations. *Acta Mater.* **52,** 2665–2671 (2004).

37. Shang, S.-L., Wang, Y., Kim, D. & Liu, Z.-K. First-principles thermodynamics from phonon and Debye model: Application to Ni and Ni3Al. *Comput. Mater. Sci.* **47,** 1040–1048 (2010).

38. Wang, Y., Liao, M., Bocklund, B. J., Gao, P., Shang, S.-L., Kim, H., Beese, A. M., Chen, L.-Q. & Liu, Z.-K. DFTTK: Density Functional Theory ToolKit for high-throughput lattice dynamics calculations. *CALPHAD* **75,** 102355 (2021).

39. DFTTK: Density Functional Theory Tool Kits. *https://www.dfttk.org/*

40. Wang, Y., Shang, S., Liu, Z.-K. & Chen, L.-Q. Mixed-space approach for calculation of vibration-induced dipole-dipole interactions. *Phys. Rev. B* **85,** 224303 (2012).

41. Wang, Y., Chen, L.-Q. & Liu, Z.-K. YPHON: A package for calculating phonons of polar materials. *Comput. Phys. Commun.* **185,** 2950–2968 (2014).

42. Wang, Y., Zhang, L. A., Shang, S. L., Liu, Z. K. & Chen, L. Q. Accurate calculations of phonon dispersion in CaF2 and CeO2. *Phys. Rev. B* **88,** 24304 (2013).

43. Mei, Z. G. First-principles Thermodynamics of Phase Transition: from Metal to Oxide. **PhD,** (The Pennsylvania State University, 2011).

44. Kaufman, L. in *Phase Stability in Metals and Alloys* (eds. Rudman, P. S., Stringer, J. S. & Jaffee, R. I.) 125–150 (McGraw-Hill, 1967).

45. Skriver, H. L. Crystal structure from one-electron theory. *Phys. Rev. B* **31,** 1909–1923 (1985).

46. Grimvall, G. Reconciling ab initio and semiempirical approaches to lattice stabilities. *Berichte Der Bunsen-Gesellschaft-Physical Chem. Chem. Phys.* **102,** 1083–1087 (1998).

47. Wang, Y., Curtarolo, S., Jiang, C., Arroyave, R., Wang, T., Ceder, G., Chen, L. Q. & Liu, Z. K. Ab initio lattice stability in comparison with CALPHAD lattice stability. *CALPHAD* **28,** 79–90 (2004).

48. Grimvall, G., Magyari-Koepe, B., Ozolins, V. & Persson, K. A. Lattice instabilities in metallic elements. *Rev. Mod. Phys.* **84,** 945–986 (2012).

49. van de Walle, A. Invited paper: Reconciling SGTE and ab initio enthalpies of the elements. *CALPHAD* **60,** 1–6 (2018).

50. van de Walle, A., Hong, Q., Kadkhodaei, S. & Sun, R. The free energy of mechanically unstable phases. *Nat. Commun.* **6,** 7559 (2015).

51. Yang, S., Wang, Y., Liu, Z. K. & Zhong, Y. Ab initio simulations on the pure Cr lattice stability at 0K: Verification with the Fe-Cr and Ni-Cr binary systems. *CALPHAD* **75,** 102359 (2021).

52. Ozolins, V. First-Principles Calculations of Free Energies of Unstable Phases: The Case of fcc W. *Phys. Rev. Lett.* **102,** 65702 (2009).

53. Kresse, G., Furthmuller, J., Furthmüller, J., Furthmueller, J., Furthmuller, J., Furthmüller, J. & Furthmü, J. Efficient iterative schemes for ab initio total-energy calculations using a plane-wave basis set. *Phys. Rev. B* **54,** 11169 (1996).

54. Levy, M. & Nagy, Á. Variational Density-Functional Theory for an Individual Excited State. *Phys. Rev. Lett.* **83,** 4361–4364 (1999).

55. Onida, G., Reining, L. & Rubio, A. Electronic excitations: density-functional versus many-body Green's-function approaches. *Rev. Mod. Phys.* **74,** 601–659 (2002).

56. Perdew, J. P. Jacob's ladder of density functional approximations for the exchange-correlation energy. in *AIP Conference Proceedings* **577,** 1–20 (AIP, 2001).

57. Tao, J., Perdew, J. P., Staroverov, V. N. & Scuseria, G. E. Climbing the Density Functional Ladder: Nonempirical Meta–Generalized Gradient Approximation Designed for Molecules and Solids. *Phys. Rev. Lett.* **91,** 146401 (2003).

58. Goerigk, L. & Grimme, S. A thorough benchmark of density functional methods for general main group thermochemistry, kinetics, and noncovalent interactions. *Phys. Chem. Chem. Phys.* **13,** 6670–6688 (2011).

59. Medvedev, M. G., Bushmarinov, I. S., Sun, J., Perdew, J. P. & Lyssenko, K. A. Density functional theory is straying from the path toward the exact functional. *Science* **355,** 49–52 (2017).

60. Kepp, K. P. Comment on "Density functional theory is straying from the path toward the exact functional". *Science* **356,** 496–496 (2017).

61. Runge, E. & Gross, E. K. U. Density-Functional Theory for Time-Dependent Systems. *Phys. Rev. Lett.* **52,** 997–1000 (1984).

62. Burke, K., Werschnik, J. & Gross, E. K. U. Time-dependent density functional theory: Past, present, and future. *J. Chem. Phys.* **123,** 062206 (2005).

63. Laurent, A. D. & Jacquemin, D. TD-DFT benchmarks: A review. *Int. J. Quantum Chem.* **113,** 2019–2039 (2013).

64. Eshuis, H. & Furche, F. A Parameter-Free Density Functional That Works for Noncovalent Interactions. *J. Phys. Chem. Lett* **2,** 983–989 (2011).

65. Chen, G. P., Voora, V. K., Agee, M. M., Balasubramani, S. G. & Furche, F. Random-Phase Approximation Methods. *Annu. Rev. Phys. Chem.* **68,** 421–445 (2017).

66. Gilbert, T. L. Hohenberg-Kohn theorem for nonlocal external potentials. *Phys. Rev. B* **12,** 2111–2120 (1975).

67. Donnelly, R. A. & Parr, R. G. Elementary properties of an energy functional of the first-order reduced density matrix. *J. Chem. Phys.* **69,** 4431–4439 (1978).

68. Zumbach, G. & Maschke, K. Density-matrix functional theory for the N -particle ground state. *J. Chem. Phys.* **82,** 5604–5607 (1985).

69. Giesbertz, K. J. H. & Ruggenthaler, M. One-body reduced density-matrix functional theory in finite basis sets at elevated temperatures. *Phys. Rep.* **806,** 1–47 (2019).

70. Anisimov, V. I., Zaanen, J. & Andersen, O. K. Band theory and Mott insulators: Hubbard U instead of Stoner I. *Phys. Rev. B* **44,** 943–954 (1991).

71. Kulik, H. J., Cococcioni, M., Scherlis, D. A. & Marzari, N. Density Functional Theory in Transition-Metal Chemistry: A Self-Consistent Hubbard U Approach. *Phys. Rev. Lett.* **97,** 103001 (2006).

72. Yamada, T. & Tohyama, T. Multipolar nematic state of nonmagnetic FeSe based on DFT+U. *Phys. Rev. B* **104,** L161110 (2021).

73. Georges, A., Kotliar, G., Krauth, W. & Rozenberg, M. J. Dynamical mean-field theory of strongly correlated fermion systems and the limit of infinite dimensions. *Rev. Mod. Phys.* **68,** 13–125 (1996).

74. Zhu, T. & Chan, G. K.-L. Ab Initio Full Cell G W + DMFT for Correlated Materials.

*Phys. Rev. X* **11,** 021006 (2021).

75. Verma, P. & Truhlar, D. G. Status and Challenges of Density Functional Theory. *Trends Chem.* **2,** 302–318 (2020).

76. Snyder, J. C., Rupp, M., Hansen, K., Müller, K.-R. & Burke, K. Finding Density Functionals with Machine Learning. *Phys. Rev. Lett.* **108,** 253002 (2012).

77. Brockherde, F., Vogt, L., Li, L., Tuckerman, M. E., Burke, K. & Müller, K.-R. Bypassing the Kohn-Sham equations with machine learning. *Nat. Commun.* **8,** 872 (2017).

78. Ellis, J. A., Fiedler, L., Popoola, G. A., Modine, N. A., Stephens, J. A., Thompson, A. P., Cangi, A. & Rajamanickam, S. Accelerating finite-temperature Kohn-Sham density functional theory with deep neural networks. *Phys. Rev. B* **104,** 035120 (2021).

79. Heyd, J., Scuseria, G. E. & Ernzerhof, M. Hybrid functionals based on a screened Coulomb potential. *J. Chem. Phys.* **118,** 8207–8215 (2003).

80. Perdew, J. P., Ruzsinszky, A., Sun, J., Nepal, N. K. & Kaplan, A. D. Interpretations of ground-state symmetry breaking and strong correlation in wavefunction and density functional theories. *Proc. Natl. Acad. Sci. U. S. A.* **118,** e2017850118 (2021).

81. Perdew, J. P., Chowdhury, S. T. U. R., Shahi, C., Kaplan, A. D., Song, D. & Bylaska, E. J. Symmetry Breaking with the SCAN Density Functional Describes Strong Correlation in the Singlet Carbon Dimer. *J. Phys. Chem. A* **127,** 384–389 (2023).

82. de Groot, S. R. & Mazur, P. *Non-equilibrium thermodynamics*. (Dover Publications, Inc. New York, 1984).

83. Kondepudi, D. & Prigogine, I. *Modern Thermodynamics: From Heat Engines to Dissipative Structures*. (John Wiley & Sons Ltd., 2015).

84. Jou, D., Casas-Vázquez, J. & Lebon, G. *Extended irreversible thermodynamics*. (Springer

Netherlands, 2010). doi:10.1007/978-90-481-3074-0

85. Lebon, G. & Jou, D. Early history of extended irreversible thermodynamics (1953-1983): An exploration beyond local equilibrium and classical transport theory. *Eur. Phys. J. H* **40,** 205–240 (2015).

86. Darling, K. A., Tschopp, M. A., VanLeeuwen, B. K., Atwater, M. A. & Liu, Z. K. Mitigating grain growth in binary nanocrystalline alloys through solute selection based on thermodynamic stability maps. *Comput. Mater. Sci.* **84,** 255–266 (2014).

87. Chen, L.-Q. Phase-field models for microstructure evolution. *Annu. Rev. Mater. Res.* **32,** 113–140 (2002).

88. Onsager, L. Reciprocal Relations in Irreversible Processes. I. *Phys. Rev.* **37,** 405–426 (1931).

89. Onsager, L. Reciprocal Relations in Irreversible Processes. II. *Phys. Rev.* **38,** 2265–2279 (1931).

90. Prigogine, I., Outer, P. & Herbo, C. Affinity and Reaction Rate Close to Equilibrium. *J. Phys. Colloid Chem.* **52,** 321–331 (1948).

91. Prigogine, I. The Equilibrium Hypothesis in Chemical Kinetics. *J. Phys. Chem.* **55,** 765–774 (1951).

92. Prigonine, I. & Résibois, P. On the kinetics of the approach to equilibrium. *Physica* **27,** 629–646 (1961).

93. Prigogine, I. & Nicolis, G. On Symmetry-Breaking Instabilities in Dissipative Systems. *J. Chem. Phys.* **46,** 3542–3550 (1967).

94. Prigogine, I. Dissipative structures, dynamics and entropy. *Int. J. Quantum Chem.* **9-S9,** 443–456 (1975).

95. Prigogine, I. Time, structure, and fluctuations. *Science* **201,** 777–785 (1978).

96. Coleman, B. D. & Truesdell, C. On the reciprocal relations of Onsager. *J. Chem. Phys.* **33,** 28–31 (1960).

97. Truesdell, C. Mechanical basis of diffusion. *J. Chem. Phys.* **37,** 2336–2344 (1962).

98. Darken, L. S. Diffusion, mobility and their interrelation through free energy in binary metallic systems. *Trans. Am. Inst. Min. Metall. Eng.* **175,** 184–201 (1948).

99. Darken, L. S. Diffusion of carbon in austenite with a discontinuity in composition. *Trans. Am. Inst. Min. Metall. Eng.* **180,** 430–438 (1949).

100. Andersson, J. & Ågren, J. Models for numerical treatment of multicomponent diffusion in simple phases. *J. Appl. Phys.* **72,** 1350–1355 (1992).

101. Kocherginsky, N. & Gruebele, M. Thermodiffusion: The physico-chemical mechanics view. *J. Chem. Phys.* **154,** 024112 (2021).

102. Liu, Z.-K. Comment on "Thermodiffusion: The physico-chemical mechanics view" [J. Chem. Phys. 154, 024112 (2021)]. *J. Chem. Phys.* **155,** 087101 (2021).

103. Ågren, J. The Onsager Reciprocity Relations Revisited. *J. Phase Equilibria Diffus.* **43,** 640–647 (2022).

104. Liu, Z. K., Wang, Y. & Shang, S.-L. Origin of negative thermal expansion phenomenon in solids. *Scr. Mater.* **65,** 664–667 (2011).

105. Liu, Z. K., Wang, Y. & Shang, S.-L. Zentropy Theory for Positive and Negative Thermal Expansion. *J. Phase Equilibria Diffus.* **43,** 598–605 (2022).

106. Liu, Z. K. Ocean of Data: Integrating first-principles calculations and CALPHAD modeling with machine learning. *J. Phase Equilibria Diffus.* **39,** 635–649 (2018).

107. Nye, J. F. *Physical Properties of Crystals: Their Representation by Tensors and Matrices*.

(Clarendon Press, 1985).

108. Onsager, L. The Motion of Ions: Principles and Concepts. *Science* **166,** 1359–1364 (1969).

109. Wang, Y., Hu, Y.-J., Bocklund, B., Shang, S.-L., Zhou, B.-C., Liu, Z.-K. & Chen, L.-Q. First-principles thermodynamic theory of Seebeck coefficients. *Phys. Rev. B* **98,** 224101 (2018).

110. Wang, Y., Chong, X., Hu, Y. J., Shang, S. L., Drymiotis, F. R., Firdosy, S. A., Star, K. E., Fleurial, J. P., Ravi, V. A., Chen, L. Q. & Liu, Z. K. An alternative approach to predict Seebeck coefficients: Application to La 3−x Te 4. *Scr. Mater.* **169,** 87–91 (2019).

111. Mantina, M., Wang, Y., Arroyave, R., Chen, L. Q., Liu, Z. K. & Wolverton, C. First-Principles Calculation of Self-Diffusion Coefficients. *Phys. Rev. Lett.* **100,** 215901 (2008).

112. Wang, Y., Xiong, Y., Yang, T., Yuan, Y., Shang, S., Liu, Z.-K., Gopalan, V., Dabo, I. & Chen, L.-Q. Thermodynamic and electron transport properties of Ca3Ru2O7 from first-principles phonon calculation. *Phys. Rev. B* **107,** 035118 (2023).

113. Bothe, D. in *Parabolic Problems: The Herbert Amann Festschrift* (eds. Escher, J., Guidotti, P., Hieber, M., Mucha, P., Prüss, J. W., Shibata, Y., Simonett, G., Walker, C. & Zajaczkowski, W.) 81–93 (Springer Basel, 2011). doi:10.1007/978-3-0348-0075-4_5

114. Allie-Ebrahim, T., Zhu, Q., Bräuer, P., Moggridge, G. D. & D'Agostino, C. Maxwell–Stefan diffusion coefficient estimation for ternary systems: an ideal ternary alcohol system. *Phys. Chem. Chem. Phys.* **19,** 16071–16077 (2017).

115. Thermo-Calc Software and Databases. Available at: http://www.thermocalc.com/.

116. CompuTherm Software and Databases. Available at: http://www.computherm.com/.

117. Zhong, W., Vanderbilt, D. & Rabe, K. M. Phase Transitions in BaTiO3 from First

Principles. *Phys. Rev. Lett.* **73,** 1861–1864 (1994).

118. Zhong, W., Vanderbilt, D. & Rabe, K. M. First-principles theory of ferroelectric phase transitions for perovskites: The case of BaTiO3. *Phys. Rev. B* **52,** 6301–6312 (1995).

119. Waghmare, U. V & Rabe, K. M. Ab initio statistical mechanics of the ferroelectric phase transition in PbTiO3. *Phys. Rev. B* **55,** 6161–6173 (1997).

120. Gordon, E., Mkhitaryan, V. V, Zhao, H., Lee, Y. & Ke, L. Magnetic interactions and spin excitations in van der Waals ferromagnet VI3. *J. Phys. D. Appl. Phys.* **54,** 464001 (2021).

121. Heine, M., Hellman, O. & Broido, D. Temperature-dependent renormalization of magnetic interactions by thermal, magnetic, and lattice disorder from first principles. *Phys. Rev. B* **103,** 184409 (2021).

122. Car, R. & Parrinello, M. Unified Approach for Molecular-Dynamics and Density-Functional Theory. *Phys. Rev. Lett.* **55,** 2471–2474 (1985).

123. Fang, H., Wang, Y., Shang, S. & Liu, Z.-K. Nature of ferroelectric-paraelectric phase transition and origin of negative thermal expansion in PbTiO3. *Phys. Rev. B* **91,** 024104 (2015).

124. Glensk, A., Grabowski, B., Hickel, T., Neugebauer, J., Neuhaus, J., Hradil, K., Petry, W. & Leitner, M. Phonon Lifetimes throughout the Brillouin Zone at Elevated Temperatures from Experiment and Ab Initio. *Phys. Rev. Lett.* **123,** 235501 (2019).

125. Troyer, M. & Wiese, U.-J. Computational Complexity and Fundamental Limitations to Fermionic Quantum Monte Carlo Simulations. *Phys. Rev. Lett.* **94,** 170201 (2005).

126. Needs, R. J., Towler, M. D., Drummond, N. D. & López Ríos, P. Continuum variational and diffusion quantum Monte Carlo calculations. *J. Phys. Condens. Matter* **22,** 023201 (2010).

127. Carlson, J., Gandolfi, S., Pederiva, F., Pieper, S. C., Schiavilla, R., Schmidt, K. E. & Wiringa, R. B. Quantum Monte Carlo methods for nuclear physics. *Rev. Mod. Phys.* **87,** 1067–1118 (2015).

128. Berg, E., Lederer, S., Schattner, Y. & Trebst, S. Monte Carlo Studies of Quantum Critical Metals. *Annu. Rev. Condens. Matter Phys.* **10,** 63–84 (2019).

129. Mondaini, R., Tarat, S. & Scalettar, R. T. Quantum critical points and the sign problem. *Science* **375,** 418–424 (2022).

130. Ceder, G. A derivation of the Ising model for the computation of phase diagrams. *Comput. Mater. Sci.* **1,** 144–150 (1993).

131. Asta, M., McCormack, R. & de Fontaine, D. Theoretical study of alloy phase stability in the Cd-Mg system. *Phys. Rev. B* **48,** 748–766 (1993).

132. van de Walle, A. & Ceder, G. The effect of lattice vibrations on substitutional alloy thermodynamics. *Rev. Mod. Phys.* **74,** 11–45 (2002).

133. van de Walle, A. Methods for First-Principles Alloy Thermodynamics. *JOM* **65,** 1523–1532 (2013).

134. Wang, Y., Hector, L. G., Zhang, H., Shang, S. L., Chen, L. Q. & Liu, Z. K. Thermodynamics of the Ce γ–α transition: Density-functional study. *Phys. Rev. B* **78,** 104113 (2008).

135. Wang, Y., Hector Jr, L. G., Zhang, H., Shang, S. L., Chen, L. Q. & Liu, Z. K. A thermodynamic framework for a system with itinerant-electron magnetism. *J. Phys. Condens. Matter* **21,** 326003 (2009).

136. Wang, Y., Shang, S. L., Zhang, H., Chen, L.-Q. & Liu, Z.-K. Thermodynamic fluctuations in magnetic states: Fe3Pt as a prototype. *Philos. Mag. Lett.* **90,** 851–859 (2010).

137. Shang, S.-L., Wang, Y. & Liu, Z.-K. Thermodynamic fluctuations between magnetic states from first-principles phonon calculations: The case of bcc Fe. *Phys. Rev. B* **82,** 014425 (2010).

138. Shang, S. L., Saal, J. E., Mei, Z. G., Wang, Y. & Liu, Z. K. Magnetic thermodynamics of fcc Ni from first-principles partition function approach. *J. Appl. Phys.* **108,** 123514 (2010).

139. Wang, Y., Shang, S. L., Hui, X. D., Chen, L. Q. & Liu, Z. K. Effects of spin structures on phonons in BaFe2As2. *Appl. Phys. Lett.* **97,** 022504 (2010).

140. Wang, Y., Saal, J. E., Shang, S. L., Hui, X. D., Chen, L. Q. & Liu, Z. K. Effects of spin structures on Fermi surface topologies in BaFe2As2. *Solid State Commun.* **151,** 272–275 (2011).

141. Wang, Y., Shang, S., Chen, L.-Q. & Liu, Z.-K. Density Functional Theory-Based Database Development and CALPHAD Automation. *JOM* **65,** 1533–1539 (2013).

142. Shang, S.-L., Wang, Y., Lindwall, G., Kelly, N. R., Anderson, T. J. & Liu, Z.-K. Cation Disorder Regulation by Microstate Configurational Entropy in Photovoltaic Absorber Materials Cu 2 ZnSn(S,Se) 4. *J. Phys. Chem. C* **118,** 24884–24889 (2014).

143. Du, J., Malyi, O. I., Shang, S.-L., Wang, Y., Zhao, X.-G., Liu, F., Zunger, A. & Liu, Z.-K. Density functional thermodynamic description of spin, phonon and displacement degrees of freedom in antiferromagnetic-to-paramagnetic phase transition in YNiO3. *Mater. Today Phys.* **27,** 100805 (2022).

144. Liu, Z. K., Shang, S.-L., Du, J. & Wang, Y. Parameter-free prediction of phase transition in PbTiO3 through combination of quantum mechanics and statistical mechanics. *Scr. Mater.* **232,** 115480 (2023).

145. Shang, S.-L., Wang, Y. & Liu, Z. K. Quantifying the degree of disorder and associated phenomena in materials through zentropy: Illustrated with Invar Fe3Pt. *Scr. Mater.* **225,** 115164 (2023).

146. Tsuchiya, T., Wentzcovitch, R. M., Da Silva, C. R. S. & De Gironcoli, S. Spin Transition in magnesiowüstite in earth's lower mantle. *Phys. Rev. Lett.* **96,** 198501 (2006).

147. Umemoto, K., Wentzcovitch, R. M., de Gironcoli, S. & Baroni, S. Order–disorder phase boundary between ice VII and VIII obtained by first principles. *Chem. Phys. Lett.* **499,** 236–240 (2010).

148. Qin, T., Wentzcovitch, R. M., Umemoto, K., Hirschmann, M. M. & Kohlstedt, D. L. Ab initio study of water speciation in forsterite: Importance of the entropic effect. *Am. Mineral.* **103,** 692–699 (2018).

149. Zhuang, J., Wang, H., Zhang, Q. & Wentzcovitch, R. M. Thermodynamic properties of ϵ - Fe with thermal electronic excitation effects on vibrational spectra. *Phys. Rev. B* **103,** 144102 (2021).

150. Umemoto, K. & Wentzcovitch, R. M. Ab initio prediction of an order-disorder transition in Mg2Ge O4: Implication for the nature of super-Earth's mantles. *Phys. Rev. Mater.* **5,** 093604 (2021).

151. Luo, C., Umemoto, K. & Wentzcovitch, R. M. Ab initio investigation of H-bond disordering in δ-AlOOH. *Phys. Rev. Res.* **4,** 023223 (2022).

152. Sun, Y., Zhuang, J. & Wentzcovitch, R. M. Thermodynamics of spin crossover in ferropericlase: an improved LDA + U sc calculation. *Electron. Struct.* **4,** 014008 (2022).

153. Wan, T., Sun, Y. & Wentzcovitch, R. M. Intermediate spin state and the B1−B2 transition in ferropericlase. *Phys. Rev. Res.* **4,** 023078 (2022).

154. Todorov, I. T., Allan, N. L., Lavrentiev, M. Y., Freeman, C. L., Mohn, C. E. & Purton, J. A. Simulation of mineral solid solutions at zero and high pressure using lattice statics, lattice dynamics and Monte Carlo methods. *J. Phys. Condens. Matter* **16,** S2751–S2770 (2004).

155. Allan, N. L., Conejeros, S., Hart, J. N. & Mohn, C. E. Energy landscapes of perfect and defective solids: from structure prediction to ion conduction. *Theor. Chem. Acc.* **140,** 151 (2021).

156. Liu, Z. K., Wang, Y. & Shang, S. Thermal Expansion Anomaly Regulated by Entropy. *Sci. Rep.* **4,** 7043 (2014).

157. Jin, J., Pak, A. J., Durumeric, A. E. P., Loose, T. D. & Voth, G. A. Bottom-up Coarse-Graining: Principles and Perspectives. *J. Chem. Theory Comput.* **18,** 5759–5791 (2022).

158. Costa, M., Goldberger, A. L. & Peng, C.-K. Multiscale Entropy Analysis of Complex Physiologic Time Series. *Phys. Rev. Lett.* **89,** 068102 (2002).

159. Wu, S.-D., Wu, C.-W., Lin, S.-G., Lee, K.-Y. & Peng, C.-K. Analysis of complex time series using refined composite multiscale entropy. *Phys. Lett. A* **378,** 1369–1374 (2014).

160. Šafránek, D., Deutsch, J. M. & Aguirre, A. Quantum coarse-grained entropy and thermodynamics. *Phys. Rev. A* **99,** 010101 (2019).

161. Liu, Z.-K. DE-SC0023185: Zentropy Theory for Transformative Functionalities of Magnetic and Superconducting Materials. Available at: https://pamspublic.science.energy.gov/WebPAMSExternal/Interface/Common/ViewPublicAbstract.aspx?rv=abfd1695-37b7-463d-9046-6eff5ac326e3&rtc=24&PRoleId=10.

162. Bekenstein, J. D. Black Holes and Entropy. *Phys. Rev. D* **7,** 2333–2346 (1973).

163. Hawking, S. W. Black holes and thermodynamics. *Phys. Rev. D* **13,** 191–197 (1976).

164. Hawking, S. W. & Page, D. N. Thermodynamics of black holes in anti-de Sitter space. *Commun. Math. Phys.* **87,** 577–588 (1983).

165. Ross, F., Hawking, S. W. & Horowitz, G. T. Entropy, area, and black hole pairs. *Phys. Rev. D* **51,** 4302–4314 (1995).

166. Bekenstein, J. D. Statistical black-hole thermodynamics. *Phys. Rev. D* **12,** 3077–3085 (1975).

167. Carlip, S. in *Physics of Black Holes* **769,** 89–123 (Springer Berlin Heidelberg, 2009).

168. Carlip, S. in *One Hundred Years of General Relativity: From Genesis and Empirical Foundations to Gravitational Waves, Cosmology and Quantum Gravity* **2,** 415–465 (World Scientific Publishing Co. Pte. Ltd., 2017).

169. Henneaux, M. & Teitelboim, C. The cosmological constant as a canonical variable. *Phys. Lett. B* **143,** 415–420 (1984).

170. Caldarelli, M. M., Cognola, G. & Klemm, D. Thermodynamics of Kerr-Newman-AdS black holes and conformal field theories. *Class. Quantum Gravity* **17,** 399–420 (2000).

171. Kastor, D., Ray, S. & Traschen, J. Enthalpy and the mechanics of AdS black holes. *Class. Quantum Gravity* **26,** 195011 (2009).

172. Kubizňák, D., Mann, R. B. & Teo, M. Black hole chemistry: thermodynamics with Lambda. *Class. Quantum Gravity* **34,** 063001 (2017).

173. Simovic, F. & Mann, R. B. Critical phenomena of charged de Sitter black holes in cavities. *Class. Quantum Gravity* **36,** 014002 (2019).

174. Johnson, C. V. Instability of super-entropic black holes in extended thermodynamics. *Mod. Phys. Lett. A* **35,** 2050098 (2020).

175. Holzhey, C., Larsen, F. & Wilczek, F. Geometric and renormalized entropy in conformal

field theory. *Nucl. Phys. B* **424,** 443–467 (1994).

176. Strominger, A. & Vafa, C. Microscopic origin of the Bekenstein-Hawking entropy. *Phys. Lett. B* **379,** 99–104 (1996).

177. Carlip, S. Logarithmic corrections to black hole entropy, from the Cardy formula. *Class. Quantum Gravity* **17,** 4175–4186 (2000).

178. Das, S., Majumdar, P. & Bhaduri, R. K. General logarithmic corrections to black-hole entropy. *Class. Quantum Gravity* **19,** 2355–2367 (2002).

179. Medved, A. J. M. & Vagenas, E. C. When conceptual worlds collide: The generalized uncertainty principle and the Bekenstein-Hawking entropy. *Phys. Rev. D* **70,** 124021 (2004).

180. Gour, G. & Medved, A. J. M. Thermal fluctuations and black-hole entropy. *Class. Quantum Gravity* **20,** 3307–3326 (2003).

181. Ghosh, A. & Mitra, P. Log correction to the black hole area law. *Phys. Rev. D* **71,** 027502 (2005).

182. Engle, J., Noui, K., Perez, A. & Pranzetti, D. The SU(2) black hole entropy revisited. *J. High Energy Phys.* **2011,** 1–30 (2011).

183. Sen, A. Logarithmic corrections to schwarzschild and other non-extremal black hole entropy in different dimensions. *J. High Energy Phys.* **2013,** 1–34 (2013).

184. Calabrese, P. & Cardy, J. Entanglement entropy and quantum field theory. *J. Stat. Mech. Theory Exp.* **2004,** P06002 (2004).

185. Coleman, P. & Schofield, A. J. Quantum criticality. **433,** 226–229 (2005).

186. Calabrese, P., Cardy, J. & Doyon, B. Entanglement entropy in extended quantum systems. *Journal of Physics A: Mathematical and Theoretical* **42,** 500301 (2009).

187. Eisert, J., Cramer, M. & Plenio, M. B. Colloquium : Area laws for the entanglement entropy. *Rev. Mod. Phys.* **82,** 277–306 (2010).

188. Calabrese, P., Cardy, J. & Tonni, E. Entanglement negativity in quantum field theory. *Phys. Rev. Lett.* **109,** 130502 (2012).

189. Alberton, O., Buchhold, M. & Diehl, S. Entanglement Transition in a Monitored Free-Fermion Chain: From Extended Criticality to Area Law. *Phys. Rev. Lett.* **126,** 170602 (2021).

190. Zhao, J., Wang, Y.-C., Yan, Z., Cheng, M. & Meng, Z. Y. Scaling of Entanglement Entropy at Deconfined Quantum Criticality. *Phys. Rev. Lett.* **128,** 010601 (2022).

191. Bird and Cage Thaumatrope. Available at: https://youtu.be/46Mlr4hvW-E.

192. Macdonald, D. D. Passivity - the key to our metals-based civilization. *Pure Appl. Chem.* **71,** 951–978 (1999).

193. Schmidhuber, J. in *Foundations of Computer Science. Lecture Notes in Computer Science* (eds. Freksa, C., Jantzen, M. & Valk, R.) 201–208 (1997). doi:10.1007/BFb0052088

194. Schmidhuber, J. Algorithmic Theories of Everything. (2000). Available at: http://arxiv.org/abs/quant-ph/0011122. (Accessed: 17th November 2022)

195. Schmidhuber, J. in *A Computable Universe* 381–398 (WORLD SCIENTIFIC, 2012). doi:10.1142/9789814374309_0020

196. Hochreiter, S. & Schmidhuber, J. Long Short-Term Memory. *Neural Comput.* **9,** 1735–1780 (1997).

197. Olson, G. B. & Liu, Z.-K. Genomic Materials Design: CALculation of PHAse Dynamics. *CALPHAD* **submitted,** (2023).

198. Saunders, N. & Miodownik, A. P. *CALPHAD (Calculation of Phase Diagrams): A*

*Comprehensive Guide*. (Pergamon, 1998).

199. Lukas, H. L., Fries, S. G. & Sundman, B. *Computational Thermodynammics: The Calphad method*. (Cambridge University Press, 2007).

200. Pelton, A. D. Phase diagrams and thermodynamic modeling of solutions. *Phase Diagrams Thermodyn. Model. Solut.* 1–383 (2018).

201. Ågren, J. CALPHAD and the materials genome A 10 year anniversary. *CALPHAD* **80,** 102532 (2023).

202. Kaufman, L. & Cohen, M. Thermodynamics and Kinetics of Martensitic Transformations. *Prog. Met. Phys.* **7,** 165–246 (1958).

203. Kaufman, L. The lattice stability of metals—I. Titanium and zirconium. *Acta Metall.* **7,** 575–587 (1959).

204. Kaufman, L. Lattice Stability of Metals II. Zink, Copper, Silver. *Bull. Am. Phys. Soc.* **4,** 181 (1959).

205. Kaufman, L., Clougherty, E. . & Weiss, R. . The lattice stability of metals—III. Iron. *Acta Metall.* **11,** 323–335 (1963).

206. Dinsdale, A. T. SGTE Data for Pure Elements. *CALPHAD* **15,** 317–425 (1991).

207. FactSage Software and Databases. Available at: https://www.factsage.com/.

208. Olson, G. B. Computational Design of Hierarchically Structured Materials. *Science* **277,** 1237–1242 (1997).

209. Kaufman, L. & Ågren, J. CALPHAD, first and second generation – Birth of the materials genome. *Scr. Mater.* **70,** 3–6 (2014).

210. Human Genome Project. (2012). Available at: https://www.genome.gov/10001772/all-about-the--human-genome-project-hgp/. (Accessed: 26th June 2018)

211. Liu, Z. K. Perspective on Materials Genome®. *Chinese Sci. Bull.* **59,** 1619–1623 (2014).

212. Kaufman, L. & Cohen, M. The martensitic transformation in the iron-nickel system-Reply. *JOM* **9,** 1314–1315 (1957).

213. Grimvall, G. Extrapolative procedures in modelling and simulations: the role of instabilities. *Sci. Model. Simulations* **15,** 5–20 (2008).

214. Kadkhodaei, S., Hong, Q. J. & Van De Walle, A. Free energy calculation of mechanically unstable but dynamically stabilized bcc titanium. *Phys. Rev. B* **95,** 064101 (2017).

215. Kadkhodaei, S. & Davariashtiyani, A. Phonon-assisted diffusion in bcc phase of titanium and zirconium from first principles. *Phys. Rev. Mater.* **4,** 043802 (2020).

216. Fattahpour, S., Davariashtiyani, A. & Kadkhodaei, S. Understanding the role of anharmonic phonons in diffusion of bcc metals. *Phys. Rev. Mater.* **6,** 023803 (2022).

217. Yang, S., Wang, Y., Liu, Z.-K. & Zhong, Y. Ab initio studies on structural and thermodynamic properties of magnetic Fe. *Comput. Mater. Sci.* **227,** 112299 (2023).

218. Krajewski, A. M., Siegel, J. W., Xu, J. & Liu, Z. K. Extensible Structure-Informed Prediction of Formation Energy with improved accuracy and usability employing neural networks. *Comput. Mater. Sci.* **208,** 111254 (2022).

219. SIPFENN: Structure-Informed Prediction of Formation Energy using Neural Networks. *https://phaseslab.com/sipfenn*

220. Choudhary, K., Zhang, Q., Reid, A. C. E., Chowdhury, S., Van Nguyen, N., Trautt, Z., Newrock, M. W., Congo, F. Y. & Tavazza, F. Computational screening of high-performance optoelectronic materials using OptB88vdW and TB-mBJ formalisms. *Sci. Data* **5,** 180082 (2018).

221. Choudhary, K., DeCost, B., Chen, C., Jain, A., Tavazza, F., Cohn, R., Park, C. W.,

Choudhary, A., Agrawal, A., Billinge, S. J. L., Holm, E., Ong, S. P. & Wolverton, C. Recent advances and applications of deep learning methods in materials science. *npj Comput. Mater.* **8,** 59 (2022).

222. Trachenko, K. & Brazhkin, V. V. Collective modes and thermodynamics of the liquid state. *Reports Prog. Phys.* **79,** 016502 (2016).

223. Wang, W. Y., Zhou, B. C., Han, J. J., Fang, H. Z., Shang, S. L., Wang, Y., Hui, X. D. & Liu, Z. K. Prediction of Diffusion Coefficients in Liquid and Solids. *Defect Diffus. Forum* **364,** 182–191 (2015).

224. Ma, J., Shang, S.-L., Kim, H. & Liu, Z.-K. An ab initio molecular dynamics exploration of associates in Ba-Bi liquid with strong ordering trends. *Acta Mater.* **190,** 81–92 (2020).

225. Hong, Q.-J. & van de Walle, A. A user guide for SLUSCHI: Solid and Liquid in Ultra Small Coexistence with Hovering Interfaces. *CALPHAD* **52,** 88–97 (2016).

226. Hong, Q.-J., Ushakov, S. V., van de Walle, A. & Navrotsky, A. Melting temperature prediction using a graph neural network model: From ancient minerals to new materials. *Proc. Natl. Acad. Sci.* **119,** e2209630119 (2022).

227. Hong, Q.-J. Melting temperature prediction via first principles and deep learning. *Comput. Mater. Sci.* **214,** 111684 (2022).

228. van de Walle, A., Asta, M. & Ceder, G. The Alloy Theoretic Automated Toolkit: A User Guide. *CALPHAD* **26,** 539–553 (2002).

229. Lu, L., Meng, X., Mao, Z. & Karniadakis, G. E. DeepXDE: A Deep Learning Library for Solving Differential Equations. *SIAM Rev.* **63,** 208–228 (2021).

230. Singh, R., Sharma, A., Singh, P., Balasubramanian, G. & Johnson, D. D. Accelerating computational modeling and design of high-entropy alloys. *Nat. Comput. Sci.* **1,** 54–61

(2021).

231. Jiang, C. & Uberuaga, B. P. Efficient Ab initio Modeling of Random Multicomponent Alloys. *Phys. Rev. Lett.* **116,** 105501 (2016).

232. Shang, S., Wang, Y. & Liu, Z.-K. ESPEI: Extensible, Self-optimizing Phase Equilibrium Infrastructure for Magnesium Alloys. in *Magnesium Technology 2010* (eds. Agnew, S. R., Neelameggham, N. R., Nyberg, E. A. & Sillekens, W. H.) 617–622 (The Minerals, Metals and Materials Society (TMS), Pittsburgh, PA, 2010).

233. Otis, R. & Liu, Z.-K. pycalphad: CALPHAD-based Computational Thermodynamics in Python. *J. Open Res. Softw.* **5,** 1 (2017).

234. PyCalphad: Python library for computational thermodynamics using the CALPHAD method. *https://pycalphad.org*

235. Bocklund, B., Otis, R., Egorov, A., Obaied, A., Roslyakova, I. & Liu, Z.-K. K. ESPEI for efficient thermodynamic database development, modification, and uncertainty quantification: application to Cu–Mg. *MRS Commun.* **9,** 618–627 (2019).

236. ESPEI: Extensible Self-optimizing Phase Equilibria Infrastructure. *https://espei.org*

237. Paz Soldan Palma, J., Gong, R., Bocklund, B. J., Otis, R., Poschmann, M., Piro, M., Wang, Y., Levitskaia, T. G., Hu, S., Kim, H., Shang, S.-L. & Liu, Z.-K. Thermodynamic modeling with uncertainty quantification using the modified quasichemical model in quadruplet approximation: Implementation into PyCalphad and ESPEI. (2022). doi:10.48550/arxiv.2204.09111

238. Li, J., Sundman, B., Winkelman, J. G. M., Vakis, A. I. & Picchioni, F. Implementation of the UNIQUAC model in the OpenCalphad software. *Fluid Phase Equilib.* **507,** 112398 (2020).

239. Egner, K., Gaube, J. & Pfennig, A. GEQUAC, an excess Gibbs energy model for simultaneous description of associating and non-associating liquid mixtures. *Berichte der Bunsengesellschaft für Phys. Chemie* **101,** 209–218 (1997).

240. Klamt, A., Krooshof, G. J. P. & Taylor, R. COSMOSPACE: Alternative to conventional activity-coefficient models. *AIChE J.* **48,** 2332–2349 (2002).

241. Peng, D.-Y. & Robinson, D. B. A New Two-Constant Equation of State. *Ind. Eng. Chem. Fundam.* **15,** 59–64 (1976).

242. Mikyška, J. & Firoozabadi, A. A new thermodynamic function for phase-splitting at constant temperature, moles, and volume. *AIChE J.* **57,** 1897–1904 (2011).

243. Li, Y., Qiao, Z., Sun, S. & Zhang, T. Thermodynamic modeling of CO2 solubility in saline water using NVT flash with the cubic-Plus-association equation of state. *Fluid Phase Equilib.* **520,** 112657 (2020).

244. Feng, X., Chen, M.-H., Wu, Y. & Sun, S. A fully explicit and unconditionally energy-stable scheme for Peng-Robinson VT flash calculation based on dynamic modeling. *J. Comput. Phys.* **463,** 111275 (2022).

245. Liu, Z.-K., Shang, S.-L. & Otis, R. A. POSE: Phase I: A Path to Sustaining a New Open-Source Ecosystem for Materials Science (OSEMatS). Available at: https://nsf.gov/awardsearch/showAward?AWD_ID=2229690.

246. Paulson, N. H., Bocklund, B. J., Otis, R. A., Liu, Z. K. & Stan, M. Quantified uncertainty in thermodynamic modeling for materials design. *Acta Mater.* **174,** 9–15 (2019).

247. Otis, R., Bocklund, B. & Liu, Z.-K. Sensitivity estimation for calculated phase equilibria. *J. Mater. Res.* **36,** 140–150 (2021).

Table 1: Physical quantities related to the first directives of molar quantities (first column) to potentials (first row), symmetrical due to the Maxwell relations [16,106].

| | $T$, Temperature | $\sigma$, Stress | $E$, Electrical field | $\mathcal{H}$, Magnetic field | $\mu_i$, Chemical potential |
|---|---|---|---|---|---|
| $S$, Entropy | Heat capacity | Piezocaloric effect | Electrocaloric effect | Magnetocaloric effect | $\frac{\partial S}{\partial \mu_k}$ |
| $\varepsilon$, Strain | Thermal expansion | Elastic compliance | Converse piezoelectricity | Piezomagnetic moduli | $\frac{\partial \varepsilon_{ij}}{\partial \mu_k}$ |
| $\theta$, Electrical displacement | Pyroelectric coefficient | Piezoelectric moduli | Permittivity | Magnetoelectric coefficient | $\frac{\partial D_i}{\partial \mu_k}$ |
| $B$, Magnetic induction | Pyromagnetic coefficient | Piezomagnetic moduli | Magnetoelectric coefficient | Permeability | $\frac{\partial B_i}{\partial \mu_k}$ |
| $N_j$, Moles | *Thermoreactivity* | *Stressoreactivity* | *Electroreactivity* | *Magnetoreactivity* | $\frac{\partial N_i}{\partial \mu_k}$, *Thermodynamic factor* |

Table 2: Cross phenomenon coefficients represented by derivatives between potentials, symmetrical due to the Maxwell relations [16,107].

| | $\boldsymbol{T}$, Temperature | $\boldsymbol{\sigma}$, Stress | $\boldsymbol{E}$, Electrical field | $\boldsymbol{\mathcal{H}}$, Magnetic field | $\boldsymbol{\mu_i}$, Chemical potential |
|---|---|---|---|---|---|
| $\boldsymbol{T}$ | 1 | $-\frac{\partial S}{\partial \varepsilon}$ | $-\frac{\partial S}{\partial \theta}$ | $-\frac{\partial S}{\partial B}$ | $-\frac{\partial S}{\partial c_i}$ Partial entropy |
| $\boldsymbol{\sigma}$ | $\frac{\partial \sigma}{\partial T}$ | 1 | $-\frac{\partial \varepsilon}{\partial \theta}$ | $-\frac{\partial \varepsilon}{\partial B}$ | $-\frac{\partial \varepsilon}{\partial c_i}$ Partial strain |
| $\boldsymbol{E}$ | $\frac{\partial E}{\partial T}$ | $\frac{\partial E}{\partial \sigma}$ | 1 | $-\frac{\partial \theta}{\partial B}$ | $-\frac{\partial \theta}{\partial c_i}$ Partial electrical displacement |
| $\boldsymbol{\mathcal{H}}$ | $\frac{\partial \mathcal{H}}{\partial T}$ | $\frac{\partial \mathcal{H}}{\partial \sigma}$ | $\frac{\partial \mathcal{H}}{\partial E}$ | 1 | $-\frac{\partial B}{\partial c_i}$ Partial magnetic induction |
| $\boldsymbol{\mu_i}$ | $\frac{\partial \mu_i}{\partial T}$ Thermodiffusio n | $\frac{\partial \mu_i}{\partial \sigma}$ Stressmigratio n | $\frac{\partial \mu_i}{\partial E}$ Electromigratio n | $\frac{\partial \mu_i}{\partial \mathcal{H}}$ Magnetomigratio n | $\frac{\partial \mu_i}{\partial \mu_j} = -\frac{\partial c_j}{\partial c_i} = \frac{\Phi_{ii}}{\Phi_{ji}}$ Crossdiffusio n |

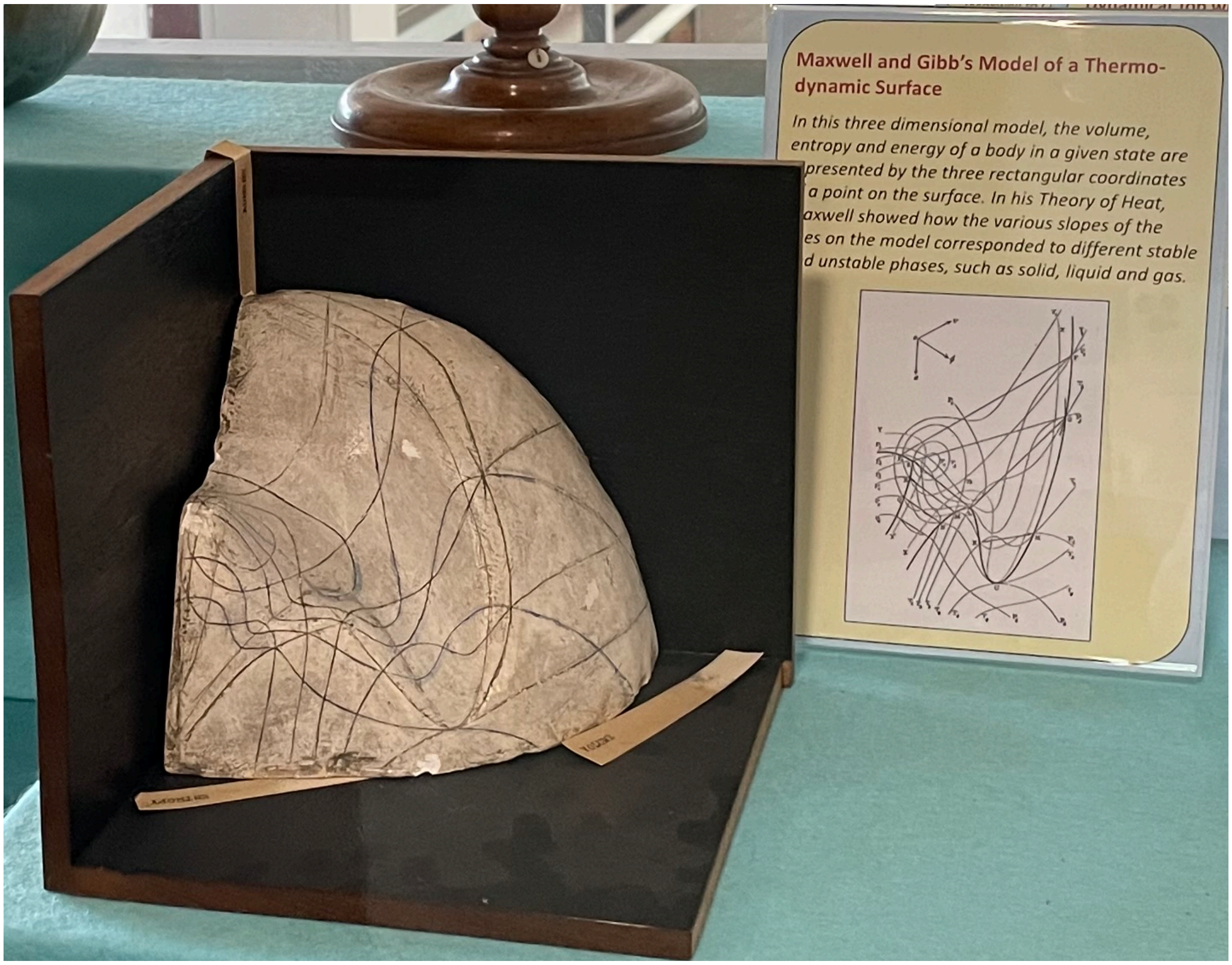
Maxwell and Gibb's Model of a Thermo-
dynamic Surface
In this three dimensional model, the volume,
entropy and energy of a body in a given state are
presented by the three rectangular coordinates
a point on the surface. In his Theory of Heat,
axwell showed how the various slopes of the
es on the model corresponded to different stable
d unstable phases, such as solid, liquid and gas.

Figure 1

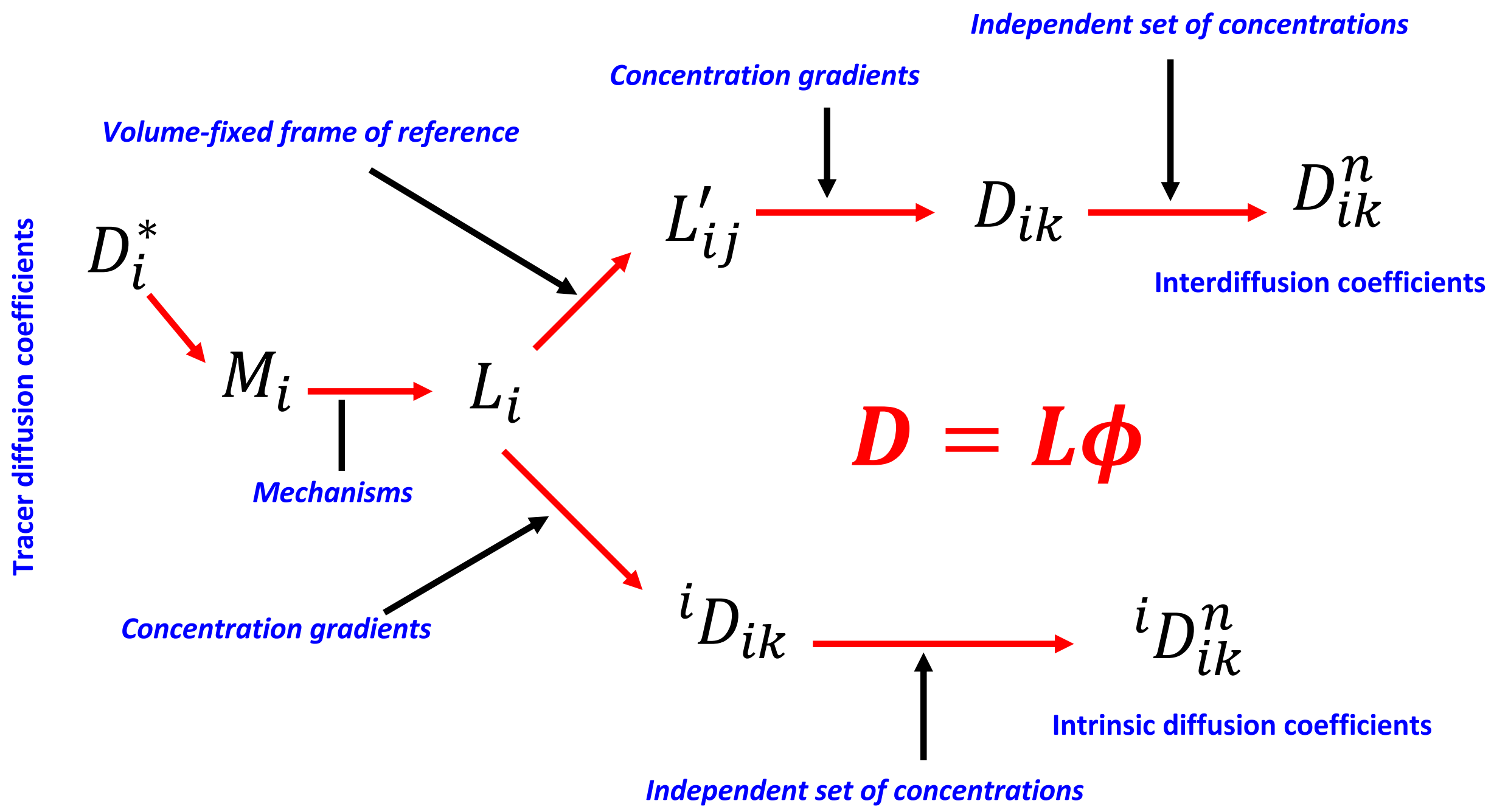

Figure 2

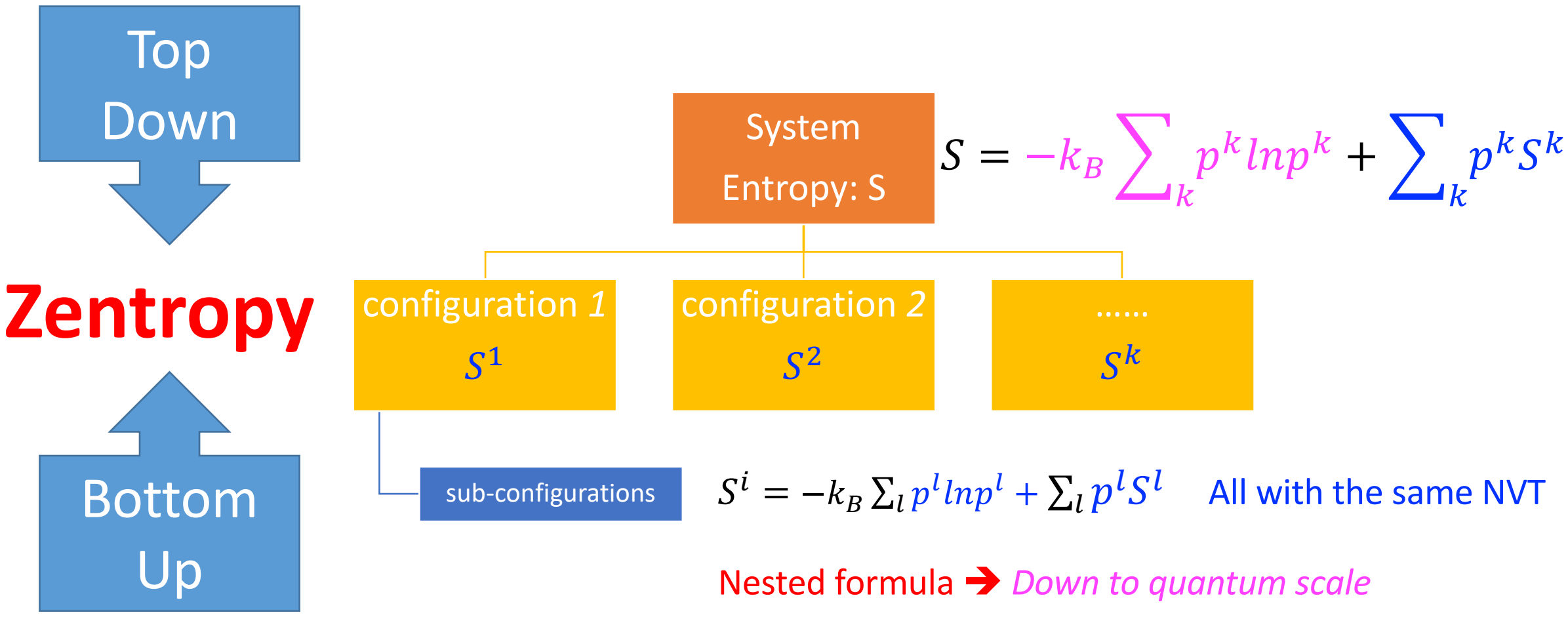

Top Down
Zentropy
Bottom Up
System Entropy: S
$S = -k_B \sum_k p^k ln p^k + \sum_k p^k S^k$
configuration 1
$S^1$
configuration 2
$S^2$
……
$S^k$
sub-configurations
$S^i = -k_B \sum_l p^l ln p^l + \sum_l p^l S^l$
All with the same NVT
Nested formula ➔ Down to quantum scale

Figure 3